\documentstyle[proceedings,numreferences,cite]{crckapb}
\begin{opening}
\title{Ab initio thermochemistry beyond chemical accuracy for
first-and second-row compounds}
\author{Jan M.L. Martin}
\institute{Department of Organic Chemistry,
Weizmann Institute of Science,
76100 Re\d{h}ovot, Israel. {\em Email:} \verb|comartin@wicc.weizmann.ac.il|
}
\end{opening}
\runningtitle{Ab initio thermochemistry beyond chemical accuracy}

\begin{document}
\newcommand{\etal}{{\em et al.}\/}
\newcommand{\IP}{inner polarization}
\newcommand{\IPF}{\IP\ function}
\newcommand{\IPFs}{\IP\ functions}
\newcommand{\jcite}[5]{(#4) #5, {\it #1} {\bf #2}, #3}
\newcommand{\auth}[2]{#2, #1, }
\newcommand{\oneauth}[2]{#2, #1}
\newcommand{\twoauth}[4]{#2, #1 and #4, #3}
\newcommand{\andauth}[2]{and #2, #1}
\newcommand{\edit}[2]{#2, #1, Ed.}
\newcommand{\et}{ and }
\newcommand{\BOOK}[4]{(#4) {\it #1}. #2, #3.}
\newcommand{\erratum}[3]{{\it erratum} (#3) {\bf #1}, #2}
\newcommand{\inbook}[5]{(#5) In {\it #1} (#2) #3, #4.}
\newcommand{\CT}[2]{(1998) #1, in {\it Computational Thermochemistry}
(Irikura, K. K. and Frurip, D. J., Eds.), ACS Symposium Series, Nr. 677, American 
Chemical Society, Washington, DC, p. #2}
\newcommand{\JCP}[4]{\jcite{J. Chem. Phys.}{#1}{#2}{#3}{#4}}
\newcommand{\jms}[4]{\jcite{J. Mol. Spectrosc.}{#1}{#2}{#3}{#4}}
\newcommand{\jmsp}[4]{\jcite{J. Mol. Spectrosc.}{#1}{#2}{#3}{#4}}
\newcommand{\theochem}[4]{\jcite{J. Mol. Struct. ({\sc theochem})}{#1}{#2}{#3}{#4}}
\newcommand{\jmstr}[4]{\jcite{J. Mol. Struct.}{#1}{#2}{#3}{#4}}
\newcommand{\cpl}[4]{\jcite{Chem. Phys. Lett.}{#1}{#2}{#3}{#4}}
\newcommand{\cp}[4]{\jcite{Chem. Phys.}{#1}{#2}{#3}{#4}}
\newcommand{\pr}[4]{\jcite{Phys. Rev.}{#1}{#2}{#3}{#4}}
\newcommand{\jpc}[4]{\jcite{J. Phys. Chem.}{#1}{#2}{#3}{#4}}
\newcommand{\jpca}[4]{\jcite{J. Phys. Chem. A}{#1}{#2}{#3}{#4}}
\newcommand{\jpcA}[4]{\jcite{J. Phys. Chem. A}{#1}{#2}{#3}{#4}}
\newcommand{\jcc}[4]{\jcite{J. Comput. Chem.}{#1}{#2}{#3}{#4}}
\newcommand{\molphys}[4]{\jcite{Mol. Phys.}{#1}{#2}{#3}{#4}}
\newcommand{\physrev}[4]{\jcite{Phys. Rev.}{#1}{#2}{#3}{#4}}
\newcommand{\mph}[4]{\jcite{Mol. Phys.}{#1}{#2}{#3}{#4}}
\newcommand{\cpc}[4]{\jcite{Comput. Phys. Commun.}{#1}{#2}{#3}{#4}}
\newcommand{\jcsfii}[4]{\jcite{J. Chem. Soc. Faraday Trans. II}{#1}{#2}{#3}{#4}}
\newcommand{\jacs}[4]{\jcite{J. Am. Chem. Soc.}{#1}{#2}{#3}{#4}}
\newcommand{\ijqcs}[4]{\jcite{Int. J. Quantum Chem. Symp.}{#1}{#2}{#3}{#4}}
\newcommand{\ijqc}[4]{\jcite{Int. J. Quantum Chem.}{#1}{#2}{#3}{#4}}
\newcommand{\spa}[4]{\jcite{Spectrochim. Acta A}{#1}{#2}{#3}{#4}}
\newcommand{\tca}[4]{\jcite{Theor. Chem. Acc.}{#1}{#2}{#3}{#4}}
\newcommand{\tcaold}[4]{\jcite{Theor. Chim. Acta}{#1}{#2}{#3}{#4}}
\newcommand{\jpcrd}[4]{\jcite{J. Phys. Chem. Ref. Data}{#1}{#2}{#3}{#4}}
\newcommand{\APJ}[4]{\jcite{Astrophys. J.}{#1}{#2}{#3}{#4}}
\newcommand{\astast}[4]{\jcite{Astron. Astrophys.}{#1}{#2}{#3}{#4}}
\newcommand{\arpc}[4]{\jcite{Ann. Rev. Phys. Chem.}{#1}{#2}{#3}{#4}}


\begin{abstract}
By judicious use of extrapolations to the 1-particle basis set
limit and $n$-particle calibration techniques, total atomization
energies of molecules with up to four heavy atoms can be obtained
with calibration accuracy (1 kJ/mol or better, on average) without
any empirical correction. For the SCF energy a 3-point 
geometric extrapolation is the method of choice. For the MP2 correlation
energy, a 2-point $A+B/(l+1/2)^3$ extrapolation is recommended, while
for CCSD and CCSD(T) correlation energies we prefer the 3-point
$A+B/(l+1/2)^C$ formula. Addition of high-exponent `inner polarization
functions' to second-row atoms is essential for reliable results. 
For the highest accuracy, accounts are required of inner-shell correlation,
atomic spin-orbit splitting, anharmonicity in the zero-point energy, and
scalar relativistic effects.
\end{abstract}

\section{Introduction and statement of the problem}

From an experimental point of view, the most fundamental thermochemical
property of a compound is its heat of formation in the gas phase. From
a theoretical point of view, it is the total atomization energy (TAE, 
$\Sigma D_0$), that is, the energy required to dissociate a ground-state
molecule into its constituent ground-state atoms in the gas phase. The 
two definitions, of course, differ merely by their choice of reference
points for the constituent elements. 

The TAE of a molecule is one of the most difficult properties, from an ab initio
perspective, to compute accurately. The use of homodesmic and isodesmic 
reactions (as shown in the contribution of Irikura\cite{Irikura} in the
present volume) may greatly accelerate convergence of the computed
result with the level of theory, but obviously presupposes that accurate
data are already available for the related compounds occurring in the
thermodynamic cycle --- a condition which is by no means always fulfilled.

The taxonomy of ab initio-based methods for theoretical thermochemistry 
(for molecular mechanics-based and semiempirical methods, see the recent
reviews of Allinger\cite{Allinger} and Thiel\cite{Thiel}, respectively)
can roughly be presented as follows:
\begin{itemize}
\item empirical corrections
\begin{itemize}
\item additive corrections
\begin{itemize}
\item pair correction schemes such as G2 theory\cite{Cur91} and its 
variants\cite{Rag98}
\item connectivity-based schemes such as Martin's 3-parameter 
correction (3PC)\cite{watoc96p,Mar98}
\item bond additivity corrections such as the BAC-MP4 
scheme\cite{Zac98}
\item atom equivalent schemes (see e.g. \cite{Irikura})
\end{itemize}
\item multiplicative corrections
\begin{itemize}
\item the PCI-X schemes of Siegbahn and coworkers\cite{PCI}
\item the SAC (scaling all correlation) schemes\cite{Gor86}
\end{itemize}
\end{itemize}
\item hybrid correction/extrapolation schemes: the CBS family of 
methods\cite{Pet96,Pet98}
\item `pure' extrapolation methods
\end{itemize}

Classifying by accuracy and applicability range, PCI-X, SAC, BAC-MP4,
and similar schemes aim at near-chemical accuracy (2--5 kcal/mol)
for large systems
--- a goal also attainable, in many cases, through modern density 
functional methods.\cite{Bec93,Bec98} G2 theory, CBS-Q, and 3PC/$spdf$
permit chemical
accuracy (about 1 kcal/mol), on average, for medium-sized molecules. 
CBS-QCI/APNO and 3PC/$spdfg$ permit mean absolute errors near 0.5 
kcal/mol, while 3PC/$spdfgh$ permits 0.24 kcal/mol (1 kJ/mol) accuracy.

The final alternative however --- which relies on no other information
than computed results for the molecule itself in a systematic sequence
of basis sets --- can reach the highest accuracies of all, 0.12 
kcal/mol (0.5 kJ/mol) on average. With the present state of computer
technology, this technique
is limited to a system with about four heavy atoms, 
although larger systems can be treated at some trade-off in accuracy.
It forms the subject of the present contribution.

Because this volume is primarily aimed at a readership of non-quantum 
chemists, we will briefly review some of the electronic structure 
methods used in this paper. Further details can be found in the 
review articles cited in the relevant sections.

\section{Major Issues}

\subsection{Treatment of dynamical correlation}

Our discussion of electron correlation methods
will be restricted to single-reference methods, i.e. methods for which 
the zero-order wave function $\psi_0$ is a single Slater determinant. We will 
in addition assume that the Hartree-Fock orbitals have been expanded 
in a finite basis set of size $N$. The Hartree-Fock equations then 
also have $N$ solutions, of which the $n$ electrons in the system fill
the occupied orbitals (for which by convention we will use indices 
$i,j,k,\ldots$). The remaining solutions constitute the space of {\em 
virtual} (unoccupied) orbitals, by convention denoted by indices 
$a,b,c,\ldots$.
The exact wave function $\psi$ can then be expanded as
\begin{eqnarray}
\psi&=&\psi_0+\sum_{i,a}{C_{ia}\psi_{i\rightarrow a}}
+\sum_{i>j,a>b}{C_{ijab}\psi_{ij\rightarrow ab}}+\ldots\nonumber\\
&\equiv&(1+\hat{C}_1+\hat{C}_2+\ldots+\hat{C}_n)\psi_0\label{fci}
\end{eqnarray}
in which the $\psi_{i\rightarrow a}$, $\psi_{ij\rightarrow ab}$, $\psi_{ijk\rightarrow abc}$, 
\ldots are singly, doubly, and triply excited configurations and the
$C_{ijk\ldots}^{abc\ldots}$ are termed configuration interaction (CI)
coefficients. (CI and related methods were very recently reviewed by
Shavitt\cite{Sha98}, where references to older reviews can also be found.)
Grouping the latter by excitation level (i.e. the number
of electrons moved from occupied to virtual orbitals), we have also
introduced the excitation operators $\hat{C}_i$ ($i$=1, 2, 3, \ldots).

A calculation in which the complete expansion, eq. (\ref{fci}), is used
is termed an FCI (full configuration interaction) calculation, and 
represents the exact solution of the nonrelativistic clamped-nuclei
Schr\"odinger equation within the given finite basis set. Unfortunately
the computational cost thereof increases factorially with the size of the
basis set and the number of electrons, 
and becomes impractical for all but the fewest-electron systems in modest basis 
sets. Where it can be done at all, even in a small basis set, it is 
an invaluable gauge for the quality of more approximate electron 
correlation methods.

An obvious approximation would be to truncate the FCI expansion at a 
given finite excitation level. This leads to limited CI methods, with 
truncation at $(1+\hat{C}_1+\hat{C}_2)$ being known as CISD (CI with 
all single and double excitations), $(1+\hat{C}_1+\hat{C}_2+\hat{C}_3)$ 
as CISDT (single, double, and triple excitations), and so forth. 
Unfortunately such methods are not {\em size-extensive}\cite{exten}, i.e. the computed
energy does not scale properly with the size of the system.\footnote{
{\em Size consistency}\cite{consis},
i.e. the property that $\lim_{r_{AB}\rightarrow\infty}{E_{AB}}=E_A+E_B$,
in addition requires correct dissociation behavior of the reference wave function.} 
This is a particularly severe disadvantage in thermochemistry since the 
inextensivity error on a typical association energy of two medium-sized
monomers $A$ and $B$ may well rival or exceed the interaction energy 
itself. 

An alternative route is to use $\psi_0$ as the 
zero-order wave function in a perturbation theoretical treatment, with
the sum of the Fock operators (of which $\psi_0$ is an eigenfunction) 
as the zero-order Hamiltonian and the difference with the true 
Hamiltonian as a small perturbation. Truncating this expansion at low
order yields the MP$n$ ($n$-th order M{\o}ller-Plesset\cite{Mol34}) or MBPT-$n$
($n$-th order many-body perturbation theory) methods. They can be 
rigorously proven\cite{Gol57} to be size
extensive at all orders. The first-order correction is 
actually included in the Hartree-Fock energy (which is why it differs 
from the sum of the orbital energies, which is the zero-order 
energy). The second-order correction is very easily and rapidly 
computed
\begin{eqnarray}
E^{(2)}&=& 
\sum_{i>j,a>b}\frac{|\left<ij||ab\right>|^2}{
\epsilon_i+\epsilon_j-\epsilon_a-\epsilon_b}
\end{eqnarray}
which explains both its popularity and its use for the basis set 
additivity steps in G2 theory\cite{Rag98} and similar methods. Note
that only double excitations enter at second order, as is the case at 
third order. Single, triple, and disconnected (see below) quadruple
excitations enter the picture at fourth order, connected quadruple 
excitations only at fifth order, and the like\cite{Kuc86}. Fifth- and even 
sixth-order methods have been implemented, but both algebraic 
complexity and mounting computational expense make higher than fourth 
orders progressively impractical (see e.g. \cite{Kuc86,Cre96a,Cre96b}).
The chief disadvantage of MP$n$ methods is that, since orbital 
energy differences appear in the denominators of the relevant energy
expressions, convergence of the MP$n$ series is very slow in the 
presence of low-lying excited states.

The third, and most satisfactory, route to an approximate solution 
is to replace eq. 
(\ref{fci}) by the equivalent ``cluster expansion''
\begin{eqnarray}
\psi&=&\exp(\hat{T}_1+\hat{T}_2+\hat{T}_3+\ldots+\hat{T}_n)\psi_0\\
\hat{T}_1\psi_0 &\equiv& \sum_{i,a}{t_{ia}\psi_{i\rightarrow a}}\\
\hat{T}_2\psi_0 &\equiv& \sum_{i>j,a>b}{t_{ijab}\psi_{ij\rightarrow ab}}\\
\hat{T}_3\psi_0 &\equiv& \sum_{i>j>k,a>b>c}{t_{ijkabc}\psi_{ijk\rightarrow abc}}\\
&\ldots&\nonumber 
\end{eqnarray}
in which the $\hat{T}_m$ 
are known as cluster operators and the 
$t_{ij\ldots ab\ldots}$ as cluster amplitudes. This leads to a 
powerful method known as coupled cluster (CC) theory.\cite{Bar89,Tay94,Sta94,Lee95,Bar95}

Obviously, a full CC expansion would not offer any material advantage 
over an FCI expansion. Truncated CC expansions however offer
two priceless advantages over their CI counterparts:
not only does the truncated CC expansion converge vastly more
rapidly than the truncated CI expansion, but it is also rigorously 
size-extensive. (See e.g. \cite{Bar89,Sta94} for proofs.)

The physical meaning of a CC expansion truncated at 
substitution order $m$ is that the wave function contains not only 
excitations up to order $m$ but also all higher excitations (up to and 
including $n$-fold), approximated by the (so-called ``disconnected'')
contribution from 
simultaneous but statistically 
independent lower-order excitations of at most order 
$m$. For instance, CCSD (coupled cluster with all single and double
substitutions,\cite{Pur82}
i.e. $\psi=\exp(\hat{T}_1+\hat{T}_2)\psi_0$) includes
such ``disconnected'' quadruple excitations ($\hat{T}_2^2/2, \hat{T}_1^2\hat{T}_2/2,
\hat{T}_1^4/24$) as arise from 
simultaneous and independent double excitations (starting at fourth 
order in MBPT) as well as those from simultaneous but independent 
single, single, and double excitations (starting at sixth order). In 
fact, CCSD contains such terms to {\em infinite} order in perturbation 
theory: what are missing are the ``connected'' quadruple excitations
(which start at fifth order) as well as disconnected terms arising from 
simultaneous single and (connected) triple excitations (also starting 
at fifth order). The CPU time requirement of CCSD, like that of CISD, scales
$\propto n^2N^4$, but it consistently 
recovers a high percentage of the exact correlation energy for most 
systems. The next step up would be
CCSDT (coupled cluster with all single, double, and triple
substitutions,\cite{ccsdt} i.e. 
$\psi=\exp(\hat{T}_1+\hat{T}_2+\hat{T}_3)\psi_0$), which yields results
exceedingly close to FCI but has a CPU time scaling $\propto n^3N^5$.
By comparison with perturbation theory, we find that the most 
important improvements over CCSD reside in fourth- and fifth-order
terms involving $\hat{T}_3$. By estimating these ``quasiperturbatively''
(i.e. using the corresponding 
perturbation theory expressions but substituting the converged $T_1$ 
and $T_2$ amplitudes for the corresponding terms in the
second- and first-order MBPT wavefunctions) we obtain the very popular
CCSD(T) method\cite{Rag89} which only has a  $\propto n^3N^4$ 
operation count (for the final (T) step) but yields results almost 
identical to CCSDT for systems where $\psi_0$ is a good zero-order
approximation\cite{Scu90}.

The QCISD and QCISD(T) methods\cite{Pop87}, which occur in G2 theory,
were originally developed as a new correlation method (``quadratic 
configuration interaction'') but can be derived by omitting certain 
terms nonlinear in $\hat{T}_1$ from 
the CCSD and CCSD(T) methods, respectively (see pp.179--181 of \cite{Tay94} 
for discussion and 
further references).

Finally, it should be pointed out that CCSD(T) energies for open-shell
systems slightly differ depending on whether an unrestricted\cite{Rag89} 
or a restricted open-shell\cite{Scu91,Wat93} 
reference was used, 
and in the latter case, on which definition for the open-shell (T)
correction was used (that of Scuseria\cite{Scu91} or that of the
Bartlett group\cite{Wat93}). Differences between the two latter definitions
are on the order of 0.1 kcal/mol or less\cite{Cra96}, but 
when considering very small differences between computed
data from different sources, care should be taken to ensure consistency.

\subsection{Static correlation and quality of the zero-order reference}

Aside from FCI which, as an exact solution, is unaffected by
the quality of $\psi_0$, all of the methods discussed above
presuppose that $\psi_0$ is a good zero-order 
approximation. Deviation from this regime (i.e. the presence of 
low-lying excited states, which leads to a situation in which
one or more excited determinants have large coefficients in $\psi$)
is known as static or nondynamical correlation. 

The quality of all nonexact single-reference electron correlation 
treatments is to a greater or lesser extent affected by nondynamical 
correlation. Hence some form of measuring its importance is essential 
in practical calculations.

One quantitative measure for the importance of static correlation is the
${\cal T}_1$ diagnostic of Lee and Taylor\cite{Lee89}, 
defined in the closed-shell case as
\begin{equation}
{\cal T}_1=\sqrt\frac{\sum_{ia}t_{ia}^{2}}{N}
\end{equation}
where $N$ is the number of electrons being correlated. 
(In the open-shell case, some double-counting needs to be avoided:
see Ref.\cite{Jay93} for details.) MP$n$, as noted before, is the most
sensitive to static correlation: experience has 
taught\cite{Lee95} that MP2 results are essentially unusable for 
${\cal T}_1$ values as low as 0.02. CCSD(T), by contrast, will produce
acceptable results for ${\cal T}_1$ values as high as 0.055, while 
QCISD(T) breaks down for somewhat lower values of ${\cal T}_1$ due to
the omission of the higher-order terms in $\hat{T}_1$.\cite{Lee90}
CCSDT is amazingly robust, yielding reliable results for, e.g., the
$X~^1\Sigma^+$ state of BN\cite{bn}, for which ${\cal T}_1$=0.08 and 
the low-lying excited state $\ldots(3\sigma)^2(4\sigma)^0(1\pi)^4(5\sigma)^2$
contributes about 30\% to the wave function. Systems with even stronger
static correlation (e.g. the Cr$_2$ molecule\cite{And96}) 
demand the use of multireference methods, which are
beyond the scope of this discussion. 

Occasionally a problem may `slip by' a  ${\cal T}_1$ test.
For instance, the 
lowest $^1\Sigma^+_g$ state of linear BNB$^+$ is almost perfectly 
biconfigurational, despite a deceptively low ${\cal T}_1$=0.040.
While the latter value does indicate strong static correlation, 
one might erroneously be led into believing CCSD(T) to be still 
applicable for this system. An alternative, but less quantitative,
criterion is inspection of the most important excited configurations 
in the converged wavefunction: yet another possibility is obtaining 
natural orbitals from a small basis set
CISD or CASSCF (complete active space 
SCF\cite{Roo87}, a multireference method) calculation and inspecting 
the natural orbital occupations. 

A related problem which should be mentioned here is symmetry breaking.
This occurs when at
geometries slightly distorted from a symmetric structure there
exists a strong near-degeneracy interaction, but the two partners of
the interaction correlate to (nearly degenerate) states of different 
symmetry at the
high-symmetry geometry. Both ${\cal T}_1$ and the SCF energy then
change drastically upon near-infinitesimal displacements from the
symmetric structure, and the potential energy surface may exhibit a 
discontinuity. This effect may be real (in
which case it is known as pseudo-Jahn-Teller effect\cite{Pea67}) or
artifactual --- in which latter case even very sophisticated electron
correlation methods based on a single-determinant SCF reference often 
fail. Aside from multireference methods, the use of Brueckner
orbitals\cite{Bru55} as the zero-order reference --- leading to the
BD\cite{BD}, BD(T)\cite{BDparT}, and BDT\cite{BDT} methods --- 
often resolves symmetry
breaking(e.g. \cite{Sta92,Bar94,Cra97}). Brueckner orbitals are defined as those
for which all $T_1$ amplitudes are identically zero, and can be alternatively
viewed as constituting the single-determinant wave function which has the
greatest overlap with the FCI wave function\cite{Kut61}. These orbitals
and the BD or BDT amplitudes are determined simultaneously in an iterative
process which will take substantially longer than a CCSD or CCSD(T) calculation,
although it has the same CPU time scaling behavior.
In the absence of 
symmetry breaking, 
BD(T) does not appear to offer significant advantages over CCSD(T)\cite{Lee92}.

\subsection{1-Particle Calibration: Quality of the
Finite Basis Set}

Perhaps the most recent and comprehensive review of basis sets is that
by Helgaker and Taylor\cite{Hel95}. We will only mention a few
salient points for the present application here.

For an atomic calculation at the SCF level, a basis set can be of 
`minimal' quality and still recover essentially the exact
SCF energy, as long as the individual functions closely mimic true
Hartree-Fock orbitals. In a molecular calculation, flexibility is
required --- which requires splitting up the valence functions ---
as well as the ability to accommodate polarization of the atomic
charge cloud in the molecular environment, which is done by 
adding higher angular momentum ($d,f,\ldots$) basis functions
(so-called {\em polarization functions}). Nevertheless, the 
basis set convergence of the SCF energy is fairly rapid compared to
the correlation energy.

In a correlated calculation on an atom, the basis set
must accommodate two important kinds of dynamical correlation effects.
The first, {\em radial correlation} (or ``in-out correlation''), 
involves the tendency of one electron to be near the nucleus when the 
other is near the periphery, or conversely. It is accommodated by 
permitting basis functions 
with extra {\em radial} nodes to mix into the wave function,
i.e. by uncontracting $s$ and $p$ functions.

The second, {\em angular correlation} (or left-right correlation),
involves the tendency of one electron to be on a different side of the
atom as the other. This is accommodated by permitting basis functions with
extra angular nodes to mix into the wave function,
i.e. by adding $d,f,g,\ldots$ functions. The convergence of 
this effect in particular is quite slow.

Note that except for the special cases of hydrogen and the alkali 
metals, the basis set extensions required for an adequate description
of radial and angular valence correlation will generally cover all the
requirements noted above for atomic SCF calculations, except for 
inner-shell polarization in second-row compounds (see next section).
The guiding principle for basis set development for high-level correlated
calculations has therefore traditionally been that a molecular basis
set should, at the very least, accommodate all basis set effects occuring
in the isolated atom.

Both main families of such basis sets in usage are based on general 
contractions\cite{Raf73}, 
i.e. all primitive Gaussians can contribute to all contracted functions.

The older of the two families are the atomic
atomic natural orbital (ANO) basis sets of Alml\"of and 
Taylor\cite{Alm87}. The starting point here are natural orbitals
obtained from an atomic CISD calculation 
in a very large primitive basis set. The natural orbitals with
the highest occupation numbers are then selected as basis functions. 
It was found that these always occur in groups of almost equal
occupation numbers: e.g., the first $f$, second $d$, and third $p$ 
function have similar natural orbital occupations.
This systematically leads, for first-row elements, to contractions like
$[4s3p2d1f]$, $[5s4p3d2f1g]$, $[6s5p4d3f2g1h]$, and so forth.
(Corresponding contractions for second-row elements are
$[5s4p2d1f]$, $[6s5p3d2f1f]$, $[7s6p4d3f2g1h]$, and the like.)

The second family, the correlation consistent ($cc$) basis sets
of Dunning\cite{Dun89} and coworkers, is establishing itself as the 
de facto standard for calibration calculations.
Dunning subjected relatively compact atomic basis 
sets to energy optimization, and considered 
the energy gain from 
adding different kinds of primitives. He then found
that these energy contributions likewise occur in groups: thus, the
energy gain from adding the first $f$, second $d$, or third $p$ 
function is similar. Again this suggests adding them in shells,
which again leads to the same typical contraction patterns as 
for their ANO counterparts. Based on the number of different
functions available for the valence orbitals, these basis sets are
known as cc-pV$n$Z (correlation consistent valence $n$-tuple zeta),
where $n$=D for double (a [3s2p1d] contraction), T for triple
(a [4s3p2d1f] contraction), Q for quadruple (a [5s4p3d2f1g] 
contraction), and 5 for quintuple zeta (a [6s5p4d3f2g1h] contraction).
For compactness, the present author and coworkers generally use the
notation V$n$Z. 

Martin\cite{Mar92} carried out a detailed comparison of computed TAE
values with equivalent ANO and $cc$ basis sets. The results were found
to be nearly identical, while the integral evaluation time for the $cc$
basis sets was considerably shorter due to their more compact primitive
size. Therefore $cc$ basis sets are more commonly used, although for 
certain other applications (like weak molecular interactions or 
electrical properties) ANO basis sets and particularly the 
``averaged ANO'' variant\cite{Wid90} may be preferable.

For the computation of electron affinities and calculations on anions 
in general, special low exponent $s$ and $p$ functions (so-called
`soft' or `diffuse' functions) are required at the SCF level (e.g.\cite{Cla83}). 
At correlated levels, the regular $s$ and $p$ functions are adequate
as radial correlation functions thereto, but angular correlation
in the tail range requires the addition of `soft' $d$, $f$, \ldots
functions.  
Kendall et al.\cite{Ken92} proposed the aug-cc-pV$n$Z basis
sets (AV$n$Z for short), 
in which the cc-pV$n$Z basis set is `augmented' with one `soft'
(or low-exponent) basis set of each angular momentum. It was 
subsequently found (e.g.\cite{hf,fno}) that these functions are 
indispensable for calculating properties such as geometries and 
harmonic frequencies of highly polar {\em neutral} molecules 
as well. It is also noteworthy\cite{hf} that including just the
soft $s$ and $p$ functions  only recovers
about half the effect.

Del Bene\cite{Del93} noted that, except in such compounds as LiH in
which hydrogen has a significant negative partial charge, omission of
the diffuse functions on H generally does not affect results. This
practice is denoted by the acronym aug$'$-cc-pV$n$Z (or A$'$V$n$Z for
short).

Some authors (e.g.\cite{Mar89}) have obtained 
excellent results for the first-row hydrides using basis sets
of only $spdf$ quality, combined with $sp$ bond functions
(i.e. basis functions centered around the bond midpoint). However, as the
extension of bond function basis sets to multiple bonds will require
$d$ bond functions, which in turn will require \cite{Mar89b} atom basis sets
of up to $spdfg$ quality to keep to keep down
basis set superposition error (BSSE, Sec. 3.4) 
to an acceptable level, the usefulness of
bond functions for our purpose appears to be somewhat limited.
Bauschlicher and Partridge\cite{BauBF} very recently compared basis set
convergence between very large atom-centered and bond function-augmented
basis sets for eight covalently bound diatomics. If the underlying 
atom-centered basis set is large enough to effectively suppress BSSE,
then bond function augmented basis sets are found not to offer a 
material improvement over purely atom-centered sets of similar size.
For weakly bound systems (like rare gas dimers), however, Tao\cite{Tao94} 
found greatly improved results upon addition of bond functions, a finding
corroborated by  Partridge and Bauschlicher\cite{BFweak}.

\subsection{1-Particle Calibration versus $n$-Particle Treatment}

Situations may arise where the electron correlation method of
choice simply cannot be used in the largest basis set that would
be preferable for an extrapolation to the infinite-basis limit.
In that case, one may need to select a less computationally demanding
method and apply the additivity approximation
\begin{eqnarray}
E({\rm Method2/Basis2})&\approx& E({\rm Method2/Basis1})
+E({\rm Method1/Basis2})\nonumber\\&-&E({\rm Method1/Basis1})
\end{eqnarray}
This presupposes, of course, that only weak coupling exists between
improvement of the $n$-particle treatment from Method1 to Method2
and enlargement of the 1-particle basis set from Basis1 to Basis2.
Needless to say, how well this assumption holds will depend to a large
extent on how different the methods and basis sets involved are, as 
well as on the system under study. For instance, in a system with 
pronounced static correlation, using MP2 to estimate basis set 
extension  effects in this manner may yield really poor results.

Several levels of such approximations can be nested, e.g.
\begin{eqnarray}
E({\rm Method3/Basis3})&\approx& E({\rm Method3/Basis1})
\nonumber\\+E({\rm Method2/Basis2})&-&E({\rm Method2/Basis1})
\nonumber\\+E({\rm Method1/Basis3})&-&E({\rm Method1/Basis2})
\end{eqnarray}
For instance,
in standard G2 theory one has Method1=MP2, Method2=MP4, Method3=QCISD(T),
and Basis1=6-311G**, Basis2=6-311+G(2df,p), \\
Basis3=6-311++G(3df,2pd).  
Below, for the atomic electron affinities,
we will consider 
Method1=CCSD(T), Method2=CCSDT, Method3=FCI, 
and Basis1=AVDZ, Basis2=AVQZ, Basis3=AV6Z.

Also, if the difference between basis sets A and B comprises two or 
more sets of basis functions that do not overlap appreciably, and cover
quite different effects and/or regions of the wave function (e.g. 
diffuse functions and core correlation functions), additivity 
approximations may be invoked. For instance, if B is the set union of B$_1$ and 
B$_2$, both of which are supersets of A, then 
\begin{equation}
E(M/B) \approx E(M/B_1) + E(M/B_2) - E(M/A)
\end{equation}
E.g. with $A$=V$n$Z, $B_1$=AV$n$Z, $B_2$=V$n$Z+IPF (inner polarization 
functions, Sec. 2.6), and $B$=AV$n$Z+IPF, this approximation was
found\cite{so2} to hold to 0.02 kcal/mol or better in CCSD(T) calculations 
on SO$_2$.

\subsection{Inner-shell Correlation}

Conventional wisdom would have it that core correlation effects will
not be important for first- and perhaps second-row compounds. The 
truth is a little more complex.

In a systematic study\cite{cc} of inner-shell correlation effects on 
atomization energies of first-row compounds, contributions as large
as 2.44 kcal/mol for C$_2$H$_2$, and 1.78 kcal/mol for CO$_2$, were
found. Clearly, one cannot afford to neglect such effects even when
striving for mere chemical accuracy.

The smaller gap between the inner-shell $(2s,2p)$ and valence $(3s,3p)$
orbitals in second-row atoms would suggest that the effects of 
inner-shell correlation would be stronger, if anything, than in the 
case of first-row compounds. While this is certainly true in terms of
the absolute correlation energy (where $(2s,2p)$ correlation may meet or 
exceed the valence contribution in importance), the differential 
contributions to the binding energy tend to be quite modest, reaching 
some 0.77 kcal/mol for SO$_2$\cite{so2} and only 0.09 kcal/mol for 
H$_2$SiO\cite{h2sio}. In fact, as found previously\cite{Gre92}, 
contributions to binding energies of silanes may actually be negative:
e.g. -0.54\cite{Gre92} or -0.56\cite{h2sio} kcal/mol in the case of
triplet silylene. 
These contributions are dwarfed by those from inner polarization (see
below).

Explicit consideration of core correlation requires the use of
special basis sets that accommodate inner-shell correlation
effects by the addition of extra radial nodes in the $s$ functions
(in practice, uncontracting the innermost $s$ function a little will
do the job) as well as `tight' or `hard' (high-exponent) extra $p$ and $d$
functions. The two main alternatives for practical calculations are
the Dunning group cc-pCV$n$Z basis sets\cite{cvtz,Pet97} (which are 
available for B through F) and the Martin-Taylor core correlation basis
sets\cite{hf}, which are available for Li through F and for Al through 
Cl. In the author's experience for first-row compounds, 
the Martin-Taylor basis set is of 
about the same quality as the cc-pCVQZ basis set, while the cc-pCVTZ 
basis set generally only recovers about 75--80\% of the inner-shell
correlation effects.

Finally, it cannot be stressed enough that including 
core correlation in basis sets not designed to handle core correlation
(such as the regular cc-pV$n$Z basis sets or the Pople basis 
sets\cite{Heh85},
which are only minimal in the core) 
will generally yield erratic core correlation contributions, and 
therefore is simply a waste of CPU time. (All electronic
structure programs presently allow for correlated calculations
with frozen core electrons.)

\subsection{Inner polarization in second-row compounds\label{ipf}}

In the course of studies (e.g., \cite{so2,sio})
on the computed geometry and vibrational 
frequencies (harmonic as well as anharmonic) of some second-row 
compounds, it was found that basis set convergence was atypically slow.
Adding a single high-exponent $d$ function to the standard Dunning 
cc-pVTZ basis set for S (which is of $[5s4p2d1f]$ quality) 
was found 
to affect the geometry in SO$_2$ by as much as 0.013 \AA\ and 1 degree, 
and the harmonic frequencies by as much as 34 cm$^{-1}$. (Similar 
effects are seen\cite{so3} in SO$_3$.) The addition of such a 
function is denoted by the ``+1'' suffix, as in VTZ+1. Likewise, the
computed total atomization energy at this level is affected by as 
much as 8 kcal/mol, an observation previously made by Bauschlicher and
Partridge\cite{Bau95} and, in passing, in the paper \cite{G1bis} on the
extension of G1  theory\cite{G1} (a predecessor of G2 theory) 
to second-row compounds.
Refs. \cite{Bau95,G1bis} both ascribed the phenomenon to hypervalence 
but, while this may play a role in the case of SO$_2$, it cannot 
account for the same phenomenon occurring in SO\cite{so2}, 
SiO\cite{sio}, and AlF\cite{sio}, none of which are hypervalent by any 
reasonable definition. The fact that a clear correlation 
exists\cite{sio}
between the polarity of the bonds and the magnitude of the effect 
supports an explanation in terms of core polarization\cite{Mul84}; the 
fact that the bulk of the effect is seen at the SCF level (as well as 
in density functional calculations for energetics\cite{Bau95}, 
geometries\cite{so3} and harmonic frequencies\cite{so3}) is 
consistent with both explanations.
Comparison of the orbitals and orbital energies from 
SCF/VTZ and SCF/VTZ+1 calculations on SO$_2$ revealed that while the 
tight $d$ function only contributes to the highest occupied valence 
orbitals, the only orbital energies seriously affected are those of the
$(2s,2p)$ like orbitals on sulfur. 

Inner-shell polarization is usually adequately accommodated by the addition of
a single tight $d$ function. The optimum values at the SCF level for 
molecules were found\cite{sio} to be surprisingly close to those of 
the tightest $d$ exponent in the Dunning cc-pV5Z basis set, which were 
therefore taken as the recommended values. In cases where the effect 
is strong, we recommend the use of even-tempered sequences 
$\alpha\beta^n$ with $\alpha$ the tighest exponent in the underlying 
basis set and $\beta$=2.5 or 3.0. Such basis sets we denote VTZ+1d,
VQZ+2d1f, and the like. 

It should be noted that since this effect is not at all present in the
separated atoms, it forms an apparent exception to the ``what is good for 
correlated atomic calculations will do for molecular ones'' rule that
generally guides basis set development. This holds true if only valence
correlation is considered: basis sets augmented for $(2s,2p)$
inner-shell correlation (such as the Martin-Taylor basis set\cite{hf})
however amply provide for inner-shell polarization, such that the above
rule prevails in a wider sense.

Infinite-basis extrapolations from a V$n$Z or AV$n$Z series tend to give 
grossly exaggerated binding energies when inner polarization is 
involved, because the $d$ and $f$ functions progressively intrude into
the `inner polarization' region as $n$ increases. The remedy obviously 
consists of adding inner polarization functions --- this should be done
in a `balanced' way since for the lower terms of the V$n$Z series, the
inner polarization functions span the same region as the tightest 
$d$ and $f$ exponents in the higher terms of the V$n$Z series.
The present author
favors the sequence VTZ+1d, VQZ+2d1f, V5Z+3d2f1g, while Bauschlicher 
and Ricca\cite{BauSO2} suggest the sequence V$n$Z+2d.

\section{Secondary Issues}

\subsection{Quality of the Zero-Point Energy}

Aside from the issue of the accuracy of the vibrational frequencies
used in the zero-point energy, one also has to contend with the
effect of anharmonicity. We will illustrate our remarks for the
case of asymmetric tops, but the conclusions are valid in general.

Correct to second order in rotation-vibration perturbation theory\cite{Pap82},
the zero-point energy of an asymmetric top is given by
\begin{equation}
{\rm ZPE} = E_0 + \frac{1}{2}\sum_i{\omega_i} 
+ \frac{1}{4}\sum_{i\geq j}{X_{ij}}\label{zpe}
\end{equation}
in which the $\omega_i$ and $X_{ij}$ are the harmonic frequencies and
first anharmonicity constants, respectively, and the $E_0$ term\cite{Tru91}
is usually very small.

Two common approximations to eq.(\ref{zpe})
are one-half the sum of the harmonic
frequencies, and one-half the sum of the vibrational fundamentals. These
approximation err on the top and bottom side, respectively: the
errors (assuming no strong Fermi resonances are present) are
\begin{eqnarray} 
{\rm ZPE} - \sum_i{\omega_i/2} &=& E_0 + \frac{1}{4}\sum_{i\geq j}{X_{ij}}\\
{\rm ZPE} - \sum_i{\nu_i/2} &=& E_0 - \frac{1}{4}\sum_{i\geq j}{X_{ij}}
- \frac{3}{4}\sum_i{X_{ii}}
\end{eqnarray}

The larger the molecule becomes, the larger these deviations will grow,
especially with molecules containing X--H bonds (which have strongly 
anharmonic stretching frequencies). The usual practice (as used, e.g., in
G2 theory) of estimating
the zero-point energy by scaling relatively low-level computed 
harmonic frequencies by a factor intended to approximate observed
fundamentals (e.g., 0.89 for HF/6-31G* frequencies\cite{Pop81}) is 
therefore not appropriate, as first suggested by Grev et al.\cite{Gre91}.
(For a sample of 14 small molecules where the exact ZPE values are known,
this procedure  was found\cite{Mar98} to result in mean and maximum
absolute errors of 0.26 and 0.72 kcal/mol,
respectively.)
Scott and Radom\cite{Sco96} (see also Ref.\cite{Pop93}) propose different
scale factors for frequencies and zero-point energies for a variety
of density functional and conventional ab initio methods.

For diatomics and some small polyatomic molecules, experimentally derived
sets of anharmonicity constants may be available, and are the
method of choice for determining zero-point energies. We follow
this approach in our highest accuracy work whenever possible
(e.g. \cite{so2,c2h4more,nh4pa,c2h4tae}). For yet others,
accurate anharmonic zero-point energies are available as 
by-products of ab initio anharmonic force field studies 
(e.g. \cite{so2,h2sio,c2h4,nh4ff,c2h2,fcch} and references therein) ---
at the levels of theory used in such studies, zero-point energies 
generally are converged to 0.1 kcal/mol or better.

One additional alternative, if both computed $\omega_i$ and observed
$\nu_i$ are available, would be to take the average of $\sum_i\omega_i/2$
and $\sum_i\nu_i/2$:
\begin{equation}
{\rm ZPE} - \frac{1}{4}\sum_i{\omega_i+\nu_i}=E_0 -  \frac{3}{8}\sum_i{X_{ii}} 
\approx 0
\end{equation}
and either estimate the diagonal stretching anharmonicities among the
$X_{ii}$ (which will be the largest) from data\cite{Hub79} for 
the corresponding diatomics or neglect that term altogether.

\subsection{Quality of the Reference Geometry}
As pointed out in, e.g., Ref.\cite{Mar98}, the leading quadratic
dependence of the total energy on displacements from the equilibrium geometry
ensures that computed thermochemical properties are fairly insensitive 
to errors in the reference geometry on the order of 0.01 \AA\ or less.
Some commonly used levels of theory for reference geometries may however
lead to much larger errors or even qualitatively incorrect geometries:
at a result, an MP2/6-31G* reference geometry for N$_2$O will cause
an error of 1.8 kcal/mol in the CCSD(T)/cc-pVTZ atomization energy.\cite{Mar98}

A particularly good compromise between accuracy and computational 
cost is offered by the B3LYP \cite{Bec93,Lee88}density functional method,
particularly with a cc-pVTZ or better basis set. (Average errors in bond
distances at the B3LYP/cc-pVTZ level are on the order of 0.003 \AA\ for
first-row compounds.\cite{dft}) For second-row atoms, the use of the
cc-pVTZ+1 basis set (see above) is desirable\cite{so3}.

\subsection{Thermal Contributions}
Except for floppy molecules, thermal contributions at room temperature
can be quite accurately evaluated using the familiar rigid 
rotor-harmonic oscillator (RRHO) approach. If data at high temperatures
are required, this approach is no longer sufficient, and an anharmonic
force field and analysis, combined with a procedure for obtaining the
rotation-vibration partition function therefrom, are required. Two
practical procedures have been proposed. The first one, due to Martin and 
coworkers\cite{Mar91,Mar92a} is based on asymptotic expansions for
the nonrigid rotor partition function inside an explicit loop over
vibration. It yields excellent results in the medium temperature range
but suffers from vibrational level series collapse above 2000 K or more.
A representative application (to FNO and ClNO) is found in Ref.\cite{fno}.

The second method, due to Topper and coworkers\cite{Top93}, is based
on Feynman path integrals, and works best in the high temperature 
limit. Therefore the two methods are complementary. 

\subsection{Basis Set Superposition Error\label{bsse}}
When one carries out a calculation on the AB diatomic using a basis
set for A and B that is incomplete (as all finite basis sets by 
definition are), the atomic energy of A in AB, and of B within AB, 
will be slightly overestimated (in absolute value) due to the fact that the basis 
functions on the other atom have become available. (It is easily 
verified that basis functions on B can be expanded as a series of higher
angular momentum functions around A.) This phenomenon is known as 
{\em basis set superposition error} (BSSE). The standard estimate is using
the Boys-Bernardi\cite{Boy70} counterpoise method:
\begin{equation}
{\rm BSSE}\approx E[A(B)]+E[B(A)]-E[A]-E[B]
\end{equation}
where $E[A(B)]$ represents the energy of A with the basis set of B 
present on a `ghost atom', and conversely for $E[B(A)]$.

While the counterpoise correction
is commonly used as a correction term for interaction energies
in weak molecular complexes, virtually no authors apply it to the 
calculation of total atomization energies, for the simple reason that 
it invariably produces {\em worse} results. In addition, the extension
of the counterpoise correction to systems with more than two fragments
is not uniquely defined\cite{Wel83,Mar89c,Par91}. 

The anomaly that neglecting BSSE would yield better results is only
an apparent one: after all, BSSE is a measure of basis set 
incompleteness --- which is precisely what we are trying to get rid of ---
but the correction has the opposite sign. For sufficiently large basis sets 
(say, of $spdfg$ quality), the NASA Ames group actually found that 
150\% of the counterpoise BSSE is a fair estimate of the remaining 
basis set incompleteness\cite{Tay92}. However, given the complications
for systems larger than diatomics, the present author prefers the use 
of extrapolation to the infinite-basis limit above such methods. 
(It goes without saying that the BSSE goes to zero at the 
infinite-basis set limit. Therefore, a sufficiently
reliable extrapolation to the infinite-basis set limits effectively
obviates the issue.)

Very recently, there has been some indication\cite{BauSO2} that inner
shell correlation contributions to TAE may exhibit (relatively speaking)
quite substantial BSSEs unless very large basis sets are used.

\subsection{Relativistic effects}

A review of relativistic quantum chemistry is beyond the scope of this
work: the reader is referred to review articles by., e.g., Dyall\cite{Dyall},
Pyykk\"o\cite{Pyk88}, Sadlej\cite{Sad95}. We will restrict ourselves
to introducing a popular approximation to relativistic effects.

Upon expanding the Dirac-Fock Hamiltonian in powers of $(v/c)^2$
($v/c$ being the fraction of the speed of light that the electron attains),
adding the Breit retardation term, and discarding higher-order terms
in $(v/c)^2$, one obtains\cite{Pyk78} the Breit-Pauli Hamiltonian:
\begin{equation}
\hat{H}=\hat{H}_{NR}-\sum_i{\frac{\nabla_i^4}{8c^2}}
+\sum_{i,A}\frac{\pi Z_A \delta(\vec{r}_i-\vec{R}_A)}{2c^2}
+\sum_i\frac{\hat{s}_i.(\nabla V.\hat{p}_i)}{2c^2}+\ldots
\end{equation}
in which $V$ is the total one-electron potential, $\delta(\vec{x})$
is a Dirac delta function, $\hat{s}$ and $\hat{p}$ are spin and momentum
operators, respectively, and two-electron components of the third
and fourth terms (which are much smaller than the corresponding
one-electron contributions) have been omitted.

The first term is the nonrelativistic Hamiltonian. The second term,
known as the mass-velocity term, arises from the relativistic 
mass increase of the electron $m=m_e/(1-(v/c)^2)$ --- in which $m_e$
represents the electron rest mass. (For the $(1s)$ orbital of a hydrogen-like
atom, $\left<v\right>/c=Z/c$.)
The third term, known as the Darwin term,
arises because\cite{Pyk78} in this approximation, the electron
is most appropriately described as a diffuse charge distribution with
dimensions on the order of $\alpha$ ($\alpha=1/c=137.037 a_0$) rather
than a point charge --- leading to reduced nuclear attraction and 
electron-electron Coulomb repulsion. (The sum of these
latter two terms is often referred to as the `scalar relativistic'
contribution.)
Finally, the fourth term represents spin-orbit coupling.

Cowan and Griffin\cite{Cow76}
suggested an approximate Hamiltonian consisting of only $\hat{H}_{NR}$
and the mass-velocity and one-electron Darwin terms --- with spin-orbit
splitting to be treated separately by perturbation theory from the 
converged wave function. (This latter approximation is only justified
if the spin-orbit splittings are much smaller in magnitude than the
electronic state splittings --- as is the case for lighter atoms.)

\subsubsection{Scalar relativistic contributions}

Martin\cite{Mar83} (no relation to the present author) went one
step further and suggested the evaluation of the Darwin and mass-velocity
terms by first-order perturbation theory. Since this approach involves
only the nonrelativistic wave function and expectation values of
one-electron operators therefrom, these relativistic corrections 
can readily be obtained from any converged {\em nonrelativistic}
Hartree-Fock or 
correlated wave function for which such expectation values can 
be evaluated, such as CISD or the averaged coupled pair functional
(ACPF) method\cite{Gda88}. 

Since the Darwin and mass-velocity (DMV) terms predominantly sample effects
near the atomic nuclei, the basis set for these types of calculations
should be flexible in the high-exponent region. Since it seems to be
obvious that inner-shell correlation would be important, a core-correlation
basis set, if necessary uncontracted in the $s$ and $p$ primitives, 
appears to be the basis set of choice.

Another technique that permits the incorporation of relativistic effects in
an otherwise nonrelativistic computational framework is the use of relativistic
effective core potentials.\cite{SBK,LANL} While this may be the only alternative
for future accurate work on, say, first-row transition metal and heavy
p-block compounds, this approach is generally not recommended for
first-and second-row compounds.

The DMV corrections usually lead to a reduction in TAE, because on average
electrons are closer to the nucleus in the separated atom than in the 
molecule. Inclusion of electron correlation usually appears to
reduce the size of the DMV terms. Since the effect will be the largest
for the innermost electrons, it is usually recommended to correlate
all electrons in calculations of the DMV contributions.

How do perturbative DMV corrections compare with 
results from more rigorous relativistic methods? 
Collins and Grev\cite{Col98} found the relativistic contribution
to the binding energy of SiH$_4$ to be $-$0.67 kcal/mol using relativistic
(Douglas-Kroll\cite{DK}) CCSD(T) in a very large basis set. At the ACPF level
with the Martin-Taylor core correlation basis set\cite{hf}, we
obtain $-$0.69 kcal/mol using 1st order Darwin and mass-velocity terms
by perturbation theory. Obviously, such excellent agreement cannot be 
automatically assumed for fourth-row, let alone fifth-row compounds.

\subsubsection{Spin-orbit coupling}

The ab initio evaluation of spin-orbit matrix elements was reviewed in
detail by Richards et al.\cite{Ric81} and recently by He{\ss} et al.\cite{Hes95}.
The most important aspect for us, however --- the atomic spin-orbit
splitting and its effect on atomization energies --- can be derived directly
from experimental data.

In a nonrelativistic calculation, the spin-orbit component states of,
for instance, B($^{2}P$), C($^{3}P$),  O($^{3}P$), and F($^{2}P$) are all 
degenerate, which of course does not hold true in Nature. This means that
any nonrelativistic calculation involving atoms with $L>0$ ground states
will intrinsically overestimate binding energies. One possible 
workaround is to adjust the experimental binding energy to obtain
``experimental nonrelativistic'' (more correctly: spin-orbit averaged)
contribution. More elegantly
the spin-orbit correction can be added to the computed binding energy.
For example, for every oxygen or sulfur atom present, the computed
TAE should be decreased 
by $[E(^3P_0)+3E(^3P_1)+5E(^3P_2)]/9-E(^3P_0)$, and for every
fluorine or chlorine atom, by $[2E(^3P_{1/2})+4E(^3P_{3/2})]/6-E(^2P_{1/2})$
(The required energy levels can be found in the JANAF tables\cite{Jan85} for the 
corresponding atoms in the gas phase.)
While these contributions are commonly neglected in more approximate
methods like G2 theory and CBS-4, one cannot do so `unpunished' in
a rigorous extrapolation calculation --- some typical contributions to
TAE for chalcogenides and halogenides of the first and second row
are 0.8 kcal/mol for F$_2$, 0.6 kcal/mol for CO$_2$, 1.0 kcal/mol
for SO$_2$, and 1.2 kcal/mol for BF$_3$. 
These contributions are clearly on the order of the 
accuracy we are trying to achieve.

\section{Extrapolation to the infinite-basis limit}

\subsection{Extrapolation of the SCF energy}
Dunning observed, in his original landmark paper on 
correlation consistent basis sets\cite{Dun89}, that the
energy gain from adding extra functions of a given angular
momentum,
as well as that from adding the first function of the next higher
angular momentum, roughly follow a geometric series. 

Feller\cite{Fel92} then noted that total energies for molecules calculated
with successive cc-pV$n$Z basis set themselves roughly followed
geometric series, and suggested the use of an expression of the form
\begin{equation}
E(n)=E_\infty+A\exp(-B n)
\end{equation}
which is itself a special case of a geometric extrapolation based on
$A+B.C^{-n}$. 

\begin{table}[h]
\caption{Comparison of performance for SCF basis set extrapolations. All 
energies in hartree\label{tab:nhf}}
\begin{tabular}{lccc}
\hline
   & numerical HF$^a$  &    Feller(Q56)$^b$   &   Schwartz5(56)$^c$ \\
\hline
Ne                     &-128.54709809   & -128.547089   & -128.547284\\
N$_2$($R$=2.068 $a_0$) &-108.9938257    & -108.993818   & -108.993988\\
BH($R$=2.336 $a_0$)    & -25.1315987    &  -25.131601   &  -25.131629\\
H$_2$($R$=1.4 $a_0$)   &  -1.13362957   &   -1.133625   &   -1.133634\\
H                      &  -0.5 exactly  &   -0.500000   &   -0.500003\\
BF($R$=2.386 $a_0$)$^d$& -124.1687792   & -124.168760   & -124.168904 \\
CO($R$=2.132 $a_0$)    & -112.790907    & -112.790890   & -112.791033 \\
\hline
\end{tabular}

(a) Refs.\cite{Mon95,Kob95}.
Bond distances $R$ taken from these references.

(b) geometric extrapolation $A+B.C^{-l}$ from SCF/cc-pVQZ, SCF/cc-pV5Z,
and SCF/cc-pV6Z energies

(c) 2-point extrapolation $A+B/(l+1/2)^5$ from SCF/cc-pV5Z
and SCF/cc-pV6Z energies

(d) aug-cc-pV$n$Z basis sets used\cite{bf3}
\end{table}

Performance of the Feller exponential 3-point extrapolation for
SCF total energies cannot be described as other than impressive.
Table \ref{tab:nhf} compares extrapolated SCF total energies with
values obtained from numerical Hartree-Fock calculations. 
The largest discrepancy, for the BF diatomic, amounts to 19 $\mu E_h$,
or 0.01 kcal/mol. A two-point $A+B/(l+1/2)^5$ formula, following a 
suggestion in Ref.\cite{Mon94}, works substantially less well.

Generally, the SCF component of atomization energies converges even
faster than these total energies, and extrapolations beyond cc-pV5Z or
aug-cc-pV5Z rarely contribute more than 0.01 kcal/mol or so. 

\subsection{Extrapolation of the valence correlation energy}

Feller originally proposed his formula as a general extrapolation for
energies, and in fact, in much of the earlier work of the Dunning group,
this formula was employed for extrapolation of the {\em total} 
CCSD(T) or MRCI energy. 

The fact that the formula is largely phenomenological and has no
physical basis would, from a pragmatic point of view, not be of
serious concern if it worked well. However, contrary to the 
SCF case, performance of the geometric extrapolation for 
correlation energies leaves something to be desired. Table 
\ref{extrap} collects error statistics for the total atomization energies
of 15 molecules for which they are very precisely (on the order of 
0.1 kcal/mol) known experimentally
(data compiled in Ref.\cite{watoc96p}, including recently revised values
for HCN\cite{hcn} and HNO\cite{hno}), 
after correction for inner-shell correlation.

\begin{table}[h]
\caption{Summary of Errors
(kcal/mol)
in Extrapolated CCSD(T) Values for TAE after Correction for Core
Correlation\label{extrap}}
\begin{tabular}{lrrrr}
\hline
&\multicolumn{2}{c}{cc-pV$n$Z}&
\multicolumn{2}{c}{aug$'$-cc-pV$n$Z}\\
& \multicolumn{2}{c}{absolute error}&\multicolumn{2}{c}{absolute error}\\
& mean & maximum & mean & maximum \\
\hline
Feller(DTQ)                 &0.72&1.86&0.66&1.50\\
Feller(TQ5)                 &0.70&1.87&0.73&1.89\\
Schwartz4(TQ)               &0.46&1.27&0.35&0.69\\
Schwartz$\alpha$(TQ5)       &0.32&0.72&0.36&1.18\\
with triple bond correction &0.22&0.64&0.23&0.78\\
Schwartz4(Q5)               &0.37&0.90&0.31&0.90\\
with triple bond correction &0.26&0.83&0.22&0.69\\
Schwartz46(TQ5)             &0.35&0.81&0.33&0.94\\
with triple bond correction &0.24&0.67&0.22&0.68\\
Separate extrapolation$^a$  &    &    &0.12&0.49\\
\hline
\end{tabular}

(a) SCF contribution Feller(TQ5); CCSD(T) valence
correlation Schwartz$\alpha$(TQ5) (see Table \ref{point12})
\end{table}

Needless to say, the conclusion that even with aug-cc-pV5Z basis sets
a mean absolute error of 0.7 kcal/mol is the best we can do seems 
rather depressing. Alternatives were therefore sought, and found.

In his pioneering contribution, Schwartz\cite{Sch63} showed that the
second-order correlation energy of a helium-like atom in a singlet
state has an asymptotic expansion of the form
\begin{equation}
\Delta E(l)=A/(l+1/2)^4+B/(l+1/2)^6+O(l^{-8})
\end{equation}
in which $\Delta E(l)$ represents the contribution of basis functions with
angular momentum $l$. Hill\cite{Hil85} then generalized this result
to a variational calculation:
\begin{equation}
\Delta E(l)=A/(l+1/2)^4+B/(l+1/2)^5+O(l^{-6})
\end{equation}

Kutzelnigg and Morgan\cite{Kut92} derived similar asymptotic expansions
of the second- and third-order MBPT energy of a two-electron atom in 
singlet as well as triplet states. For the singlet, they found the expansion
\begin{equation}
\Delta E(l)=A/(l+1/2)^4+B/(l+1/2)^5+C/(l+1/2)^6+O(l^{-7})\label{schwartz}
\end{equation}
(where the $l^{-5}$ term has no second-order contribution)
while for the triplet, the expansion starts two orders later, at $(l+1/2)^{-6}$.
As pointed out in Ref.\cite{Kut92}, this result can be generalized to the
second-and third-order energies of many-electron atoms having an asymptotic
correlation energy expansion of the form eq. (\ref{schwartz}). 

If so, the error for a calculation in a basis set truncated at 
angular momentum $L$ is given by
\begin{eqnarray}
E_\infty-E(L)&=&\sum_{l=L+1}^\infty\left[\frac{A}{(l+1/2)^4}
+\frac{B}{(l+1/2)^5}+\ldots\right]\\
&=&\frac{A\psi^{(3)}(L+3/2)}{6}+
\frac{B\psi^{(4)}(L+3/2)}{24}+\ldots
\end{eqnarray}
where $\psi^{(n)}(x)$ represents the polygamma function\cite{Abr72}
of order $n$. Its asymptotic expansion has the leading terms
\begin{eqnarray}
\psi^{(n)}(x)&=&(-1)^{n-1}[\frac{(n-1)!}{x^n}+\frac{n!}{2x^{n+1}}+O(x^{-n-2})]\nonumber\\
&=& (-1)^{n-1}\frac{(n-1)!}{(x-1/2)^n}+O(x^{-n-2})
\end{eqnarray}
Hence
\begin{equation}
E(L)=E_\infty-\frac{A(L+1)^{-3}}{3}+
\frac{B(L+1)^{-4}}{4}+O(L^{-5})\label{xx}
\end{equation}
Assuming that higher orders in perturbation theory would behave similarly,
the idea of carrying out successive (say, CCSD(T)) calculations in 
completely saturated basis sets up to given angular momenta $L_1$, $L_2$,
$L_3$, followed by an extrapolation, then naturally suggests itself. 
In practice complete saturation of a basis set up to a given angular momentum
$L$ is not necessarily the most computationally efficient alternative; the
next best solution would be to use a sequence of basis sets which are 
balanced in their quality for radial and angular correlation, such as
the ANO or correlation consistent basis sets. 

If we identify $L$ with the $n$ in the cc-pV$n$Z basis sets, an ambiguity
arises, in that the highest angular momentum present in the cc-pV$n$Z 
basis set is $n$ for first-and second-row atoms, and $n-1$ for hydrogen
and helium. As a compromise, we proposed\cite{l4} an extrapolation
in terms of inverse powers of $L+1/2$. 

Extending this approach to molecular calculations involves not so much
a leap of faith as the suggestion that molecular correlation effects 
would be predominantly atomic in character. We will introduce the
following notations for two-point extrapolations to cc-pV$l$Z and 
cc-pV$m$Z energies:

\begin{tabular}{lr}
Schwartz3($kl$) & $A+B/(l+1/2)^3$\\
Schwartz4($kl$) & $A+B/(l+1/2)^4$\\
\end{tabular}

and for three-point extrapolations to cc-pV$l$Z, cc-pV$m$Z, and 
cc-pV$n$Z energies:

\begin{tabular}{lr}
Schwartz46($klm$) & $A+B/(l+1/2)^4+C/(l+1/2)^6$\\
Schwartz$\alpha$($klm$) & $A+B/(l+1/2)^\alpha$\\
\end{tabular}

and so forth. (Note that the parameters in Schwartz$\alpha$ have to be 
determined iteratively, while the others can be found by solving a 
2$\times$2 or $3\times3$ linear system.)

\begin{table}[t]
\caption{Comparison of extrapolated and essentially exact MP2 
valence correlation energies ($E_h$)\label{mp2}}

\begin{tabular}{lccccc}
\hline
   &  MP2-R12$^a$ &   Schwartz$\alpha$(Q56) &$\alpha$&  Schwartz3(56) &
   Feller(456)\\
\hline
H$_2$O$^b$ &  -0.30053 & -0.29991 &  3.44 &  -0.30069 & -0.298325\\
Ne     &  -0.3202  & -0.31985 &  3.21 &  -0.32047 &  -0.317078\\
N$_2$  &  -0.42037 & -0.41928 &  3.44 &  -0.42027 &  -0.417277\\
HF     &  -0.3198  & -0.31931 &  3.31 &  -0.32003 & -0.317190\\
\hline
\end{tabular}

(a) MP2 in basis set with explicit interelectronic bond distances\cite{Klo95}

(b) AV$n$Z basis set on O, V$n$Z basis set on H

\end{table}

Let us first consider the MP2 energy. 
Klopper\cite{Klo95} obtained what are considered essentially exact
MP2 correlation energies for N$_2$, H$_2$O, Ne, and HF using
an explicitly correlated method. As seen below in Table \ref{mp2},
a Schwartz3(56) extrapolation to MP2/AV5Z and MP2/AV6Z correlation
energies yields values in excellent agreement with the MP2-R12 
results: deviations are
-0.27 m$E_h$ for Ne, -0.25 m$E_h$ for HF, +0.10 m$E_h$
for N$_2$, and -0.14 m$E_h$ for H$_2$O, leading to
a mean absolute deviation of 0.19 m$E_h$. (Wilson and Dunning\cite{Wil97}
found similar results with V$n$Z basis sets.) A Schwartz$\alpha$(Q56) 
extrapolation to MP2/AVQZ, MP2/AV5Z, and MP2/AV6Z actually results in
less good agreement (mean absolute deviation of 0.6 m$E_h$). 
This clearly suggests that
Schwartz3 is the extrapolation of choice for large basis set MP2
calculations, as well as that convergence of the MP2 energy for
aug-cc-pV5Z and larger basis sets is almost entirely dominated by
the leading Schwartz expansion term.
Varying the $\alpha$ exponent and adding MP2/AVQZ results does not
result in an improvement: it appears that for basis sets this size,
the $(l+1/2)^{-3}$ term dominates basis set convergence. By contrast,
Feller(456) undershoots the MP2-R12 results by as much as 3 millihartree.

\begin{table}[h]
\caption{Comparison of extrapolated and essentially exact CCSD and CCSD(T) 
valence correlation energies ($E_h$)\label{ccsd}}

\begin{tabular}{lccccc}
\hline
   &  CCSD-R12$^a$ &   Schwartz$\alpha$(Q56) &$\alpha$&  Schwartz3(56) &
   Feller(456)\\
\hline
H$_2$O$^b$  & -0.29753 & -0.29755 & 3.96  &  -0.29853  &      -0.29668\\
Ne$^b$   & -0.31542 & -0.31519 & 3.64  &  -0.31650  &      -0.31343\\
F-   & -0.32262 & -0.32207 & 3.76  &  -0.32326  &      -0.32076\\
HF   & -0.31391 & -0.31359 & 3.77  &  -0.31472  &      -0.31234\\
\hline
   &  CCSD(T)-R12$^a$ &   Schwartz$\alpha$(Q56) &$\alpha$&  Schwartz3(56) &
   Feller(456)\\
\hline
H$_2$O$^c$  & -0.30737 & -0.30734 & 3.97  &  -0.30842  &       -0.30648\\
Ne$^c$   & -0.32167 & -0.32165 & 3.67  &  -0.32305  &       -0.31986\\
HF   & -0.32245 & -0.32238 & 3.80  &  -0.32360  &       -0.32110\\
F-   & -0.33427 & -0.33402 & 3.80  &  -0.33532  &       -0.33266 \\
\hline
\end{tabular}

(a) from CCSD-R12 and CCSD(T)-R12 results in Ref.\cite{Mul97} derived as
CCSD-R12(valence)/X + CCSD-R12(all)/Y - CCSD-R12(valence)/X
(with X being their smaller and Y their bigger basis set). 

(b) more recent results\cite{Hal98}: -0.297527 (H$_2$O) and 
-0.315523 (Ne) $mE_h$

(c) more recent results\cite{Hal98}: -0.307211 (H$_2$O) and
-0.321882 (Ne) $mE_h$

\end{table}

Very recently, M\"uller, Kutzelnigg, and Noga (MKN)\cite{Mul97} (see Table 4)
published CCSD-R12 and CCSD(T)-R12 studies on a number of closed-shell
ten-electron systems, including F$^-$, HF, Ne, and H$_2$O. Some further
results of this type are available from the work of Halkier et 
al.\cite{Hal98}. MKN
quote all-electron results with two basis sets which we will denote
A and B, but valence-only results only with the smaller of the two basis sets,
A. 
Since the main improvement in basis B over basis A is in the valence
region and, in Ref.\cite{Klo95}, a basis set equivalent to A
appeared to yield inner-shell pair energies essentially equivalent to
exact solution for Ne, we would argue that the main deficiency in A
will be for the valence region and not for the inner-shell region.
Therefore, the exact valence only CCSD(T) energy is expected to lie close
to valence(A)+all(B)-all(A).

Again, Feller(456) undershoots the CCSD-R12 and CCSD(T)-R12 results by
several millihartree. Schwartz3(56) appears to overshoot the energies,
while Schwartz$\alpha$(456) appears to be in close agreement. It should
be noted that the exponent $\alpha$ here systematically favors values
significantly higher than 3, in fact centering around 4. (This
tendency is what led, in our first paper\cite{l4} on these extrapolations, to
the suggestion of Schwartz4 and Schwartz46 as extrapolation formulas.)
While Helgaker and coworkers\cite{Hal98,Hel97} advocate the use of a fixed
exponent of 3 for CCSD and CCSD(T) correlation energies as well, we
would argue that
the difference with the convergence behavior at the MP2 level reflects the
importance of higher-order terms in eq.(\ref{schwartz}) in methods
that include higher-order MBPT terms (such as CCSD and CCSD(T), both of
which include the complete third-order energy, as well as important
subclasses of excitations to infinite order). This is also consistent with
the observation of the present author\cite{coupling} who found that the
basis set increment ratio
\begin{equation}
\frac{
\hbox{TAE[MP2/AV$n$Z] -- TAE[MP2/AV$(n-1)$Z]}}{
\hbox{TAE[CCSD/AV$n$Z] -- TAE[CCSD/AV$(n-1)$Z]}}
\end{equation}
becomes progressively larger as $n$ increases, and exceeds unity
for $n$=4 and upwards.

\begin{table}[t]
\caption{Computed (CCSD(T)), extrapolated, and observed total atomization energies and
auxiliary quantities. All units are kcal/mol except
$\alpha$, which is dimensionless.\label{point12}}
\begin{tabular}{lccccccc}
\hline
 &Observed&\multicolumn{4}{c}{Extrapolated}&$\alpha$ & core corr\\
 & (a)    & total & (b) & SCF & val.corr. &  & (a)\\
\hline
HNO        &  205.64(6)  & 205.30 & 205.67 & 85.44  & 119.38 & 3.89 & 0.48\\
CO$_2$     &  389.68(6)  & 389.75 & 389.75 & 258.08 & 129.89 & 3.91 & 1.78\\
CO         &  259.58(12) & 259.56 & 259.56 & 181.59 &  77.01 & 3.69 & 0.96\\
F$_2$      &  39.01(10)  & 38.29  & 38.29  & -31.07 &  69.43 & 4.26 &-0.07\\
N$_2$      &  228.42(3)  & 228.16 & 228.53 & 119.71 & 107.59 & 3.52 & 0.85\\
N$_2$O     &  270.60(10) & 269.73 & 270.23 &  95.15 & 173.32 & 3.93 & 1.26\\
C$_2$H$_2$ &  405.53(24) & 405.04 & 405.04 & 299.93 & 102.67 & 4.37 & 2.44\\
CH$_4$     &  420.23(14) & 420.18 & 420.18 & 331.60 &  87.33 & 4.55 & 1.25\\
H$_2$CO    &  374.09(16) & 374.33 & 374.33 & 264.86 & 108.15 & 4.13 & 1.32\\
H$_2$O     &  232.83(2)  & 232.83 & 232.83 & 160.03 &  72.41 & 4.66 & 0.38\\
H$_2$      &  109.48(0)  & 109.48 & 109.48 &  83.86 &  25.62 & 4.31 & 0.00\\
HCN        &  313.27(25) & 312.96 & 313.33 & 204.42 & 106.86 & 3.94 & 1.67\\
HF         &  141.57(17) & 141.54 & 141.54 & 100.04 &  41.32 & 5.38 & 0.18\\
NH$_3$     &  298.06(10) & 297.77 & 298.15 & 203.31 &  93.79 & 4.44 & 0.66\\
C$_2$H$_4$ &  563.68(29) & 563.77 & 563.77 & 435.11 & 126.30 & 4.25 & 2.36\\
\hline
\multicolumn{2}{c}{mean abs. err. w/o F$_2$}  & 0.23 & 0.12 &  &  &  & \\
\hline
\end{tabular}

(a) see Ref.\cite{c2h4tae} for detailed references

(b) using the additional correction $\sum_i{(\hbox{bond
order})_i}\times$0.126 kcal/mol, where $i$ runs over all bonds with at
least one N atom

\end{table}

Following the adage ``the proof of the pudding is in the eating'', 
we considered\cite{c2h4tae} 
separate extrapolation of the 
CCSD(T)/A$'$V$n$Z ($n$=T, Q, 5) valence correlation component of TAE 
using Schwartz$\alpha$(TQ5), and of the SCF component
to Schwartz5(Q5) (or Feller(TQ5) --- the results differ negligibly), for our
15 reference molecules.
Agreement with experiment (Table \ref{point12})
speaks for itself, with mean and maximum
absolute errors of 0.23 and 0.88 kcal/mol. If an additional small 
correction is introduced\cite{c2h4tae} for the especially slow basis set
convergence in nitrogen compounds (0.126 kcal/mol per bond order 
involving N), mean and maximum errors can be further brought down
to 0.12 and 0.49 kcal/mol, respectively --- benchmark quality by any
reasonable standard. The same methodology was
also applied to the first-row hydrides and hydride radicals AH$_n$ 
\cite{hydride}, and some variants were considered and extensively
tested in Refs.\cite{Fel98,May98}.

Finally, 
it is perhaps worth mentioning that the convergence of the {\em sum}
of SCF and correlation energies for relatively small basis sets 
(particularly if the total energy, rather than TAE, is considered) would
be dominated to a substantial extent by the SCF convergence behavior,
which would lead to the erroneous conclusion that overall convergence
behavior is best described by an exponential series.

\subsection{Individual or global extrapolation?}

For the two-point $A+B/(l+1/2)^n$ extrapolations ($n$ fixed), 
it is easily seen that extrapolation on individual energies
or any reaction energy yields identical results. With the other
extrapolations, this equality does not hold.
In most cases, the final result for total atomization energies
will only differ by about 0.1 kcal/mol between the two approaches.
Two observations are relevant here.

First of all, as seen for the example of the ten-electron
hydrides in Table \ref{percent} below, the correlation component to
atomization energies appears to converge faster than that of the
constituent total energies. For instance, while the percentage of 
the valence correlation energy recovered by the AV5Z basis set 
varies from 99.0 \% for CH$_4$ to 97.8\% in HF, consistently 99.5\% to
99.7\% is recovered for the 
valence correlation component to the dissociation energy, which suggests
a significant cancellation of correlation effects between atom and 
molecule.

\begin{table}[h]
\caption{Convergence of CCSD(T) valence correlation component ($E_h$)
of the total energies of A, AH$_n$ (A=B--F), and the total atomization
energy of AH$_n$\label{percent}}
\begin{tabular}{lrrrrrr}
\hline
 & Extrapolated & $\alpha$ & \multicolumn{4}{c}{Percentage recovered at}\\
& valence corr.  &  & \multicolumn{4}{c}{CCSD(T)/aug$'$-cc-pV$n$Z level}\\
 & energy ($E_h$) &          & $n$=D & $n$=T & $n$=Q & $n$=5 \\
\hline
B & -0.072849 & 4.250 & 88.1 & 96.4 & 98.8 & 99.5 \\
C & -0.100642 & 3.841 & 81.2 & 94.3 & 97.8 & 99.0 \\
N & -0.129075 & 3.634 & 75.1 & 92.3 & 96.9 & 98.5 \\
O & -0.192694 & 3.421 & 69.9 & 89.4 & 95.5 & 97.7 \\
F & -0.257675 & 3.277 & 67.7 & 87.6 & 94.6 & 97.2 \\
Ne & -0.323458 & 3.194 & 65.8 & 86.4 & 93.9 & 96.8 \\
\hline
BH$_3$ & -0.145407 & 4.698 & 81.4 & 94.9 & 98.4 & 99.4 \\
CH$_4$ & -0.239761 & 4.262 & 80.9 & 94.4 & 98.1 & 99.2 \\
NH$_3$ & -0.278460 & 4.034 & 78.3 & 93.1 & 97.5 & 98.9 \\
H$_2$O & -0.307884 & 3.743 & 74.9 & 91.3 & 96.6 & 98.4 \\
HF & -0.323184 & 3.493 & 71.0 & 89.1 & 95.5 & 97.8 \\
Ne & -0.323458 & 3.194 & 65.8 & 86.4 & 93.9 & 96.8 \\
\hline
BH$_3$$\rightarrow$B+3H  & -0.072570 & 4.94  & 74.5 & 93.4 & 98.1 & 99.3 \\
CH$_4$$\rightarrow$C+4H  & -0.138758 & 4.55  & 81.0 & 94.8 & 98.5 & 99.6 \\
NH$_3$$\rightarrow$N+3H  & -0.148944 & 4.44  & 81.3 & 94.1 & 98.3 & 99.5 \\
H$_2$O$\rightarrow$O+2H  & -0.115048 & 4.66  & 83.6 & 94.4 & 98.5 & 99.6 \\
HF$\rightarrow$F+H       & -0.065716 & 5.38  & 85.6 & 94.9 & 98.8 & 99.7 \\
\hline
\end{tabular}
\end{table}

Secondly, it is easily seen from considering the difference of two
asymptotic series (for $E_{\rm corr}$(X) and $E_{\rm corr}$(Y)) in $(l+1/2)$
\begin{equation}
A_x-A_y+(B_x-B_y)/(l+1/2)^3+(C_x-C_y)/(l+1/2)^4
+\ldots
\end{equation}
that situations may arise in which the coefficients of the {\em difference}
do not decay as fast with increasing $l$ as one might like,
e.g. if X and Y are close in energy to begin with.
Under such circumstances, a three-point extrapolation 
of the form $A+B/(l+1/2)^C$ may not behave well numerically, and 
extrapolation on the individual energies may be preferable. 
(We found this to be the case, for instance, with electron affinities.)

As a rule, the present author favors extrapolation on the energy difference
when the latter is fairly large (e.g. total atomization energies), and
extrapolation on the constituent total energies when small energy
differences have to be determined (e.g. electron affinities, conformational
energy differences).

\section{Case studies}

\subsection{Electron affinities of the first-row atoms}

Electron affinities have in the past been notorious as a `tough nut
to crack' computationally. On the one hand, 
wave functions for anions extend fairly far in
space (requiring accomodation thereof in the basis set by the addition
of `diffuse functions'). On the other hand, electron affinities involve
small differences in energy between systems with different numbers of
electrons --- a balanced description of which is a very taxing test
for electron correlation methods.

As a result, the electron affinities of the first-row atoms have 
traditionally been used as a benchmark for computational methods
(e.g.\cite{Rag85,Ken92,Nor91,Bau86a,Bau86b})
not only because of the small size of the system but because the
corresponding experimental quantities are accurately known.\cite{CRC78}

This problem is a good illustration of the issues that enter
if one wants to carry out a calculation to the very highest accuracy.

Since even for atoms, full configuration interaction is not an option
with sufficiently large basis sets, we will use CCSD(T) as our `baseline'
electron correlation method. Our computed results are given in Table
\ref{ea}.

\begin{table}[h]
\caption{Computed (this work) and observed electron affinities (eV) of
the first-row atoms\label{ea}}
{\scriptsize
\begin{tabular}{lcccccccc}
\hline
  & SCF  & CCSD(T)&  core &  \multicolumn{2}{c}{Relativistics} & FCI corr. & best  &  Expt.\cite{CRC78}\\
  &      & val.corr.&  corr.&  spin-orbit  &   scalar & correction & calc.& \\
\hline
H &-0.3288 &1.0821 & 0     & 0     & -4$\times10^{-5}$ & 0$^a$  &   0.7533 &0.75420(2)\\
B &-0.2675 &0.5245 & 0.0043&-0.0005& -0.0013 &  0.0191$^b$&0.2786 &0.277(10)\\
C & 0.5483 &0.7001 & 0.0072&-0.0037& -0.0028 & 0.0140$^c$&1.2631 &1.2629(3)\\
N &\multicolumn{8}{c}{anion not bound}\\
O & -0.5390 &1.9936 & 0.0017&-0.0023 & -0.0059 & 0.0114$^d$&1.4595 &1.461122(3)\\
F & 1.3073 &2.1175  &0.0043&-0.0167 & -0.0093 & -0.0004$^d$& 3.4027 & 3.401190(4)\\
\hline
\\
\end{tabular}
}

SCF: exponential extrapolation from SCF/AVQZ, SCF/AV5Z, and SCF/AV6Z

valence correlation: $A+B/(l+1/2)^C$ extrapolation on correlation energy from
CCSD(T)/AVQZ, CCSD(T)/AV5Z, and CCSD(T)/AV6Z

spin-orbit coupling: from experimental fine structure\cite{Jan85}

scalar relativistics: Darwin and mass-velocity terms by perturbation
from ACPF/ACVQZ(uncontracted)

core correlation: CCSD(T)/ACV5Z(all)-CCSD(T)/ACV5Z(valence)

FCI correction: difference between CCSD(T) and FCI. See following footnotes:

(a) CCSD(T) is exact for a two-electron system

(b) FCI/AVQZ $-$ CCSD(T)/AVQZ

(c) FCI/AVTZ $-$ CCSD(T)/AVQZ

(d) CCSDT/AVQZ $-$ CCSD(T)/AVQZ + FCI/AVDZ $-$ CCSDT/AVDZ

\end{table}

For the SCF and valence correlation contributions, we have carried out
CCSD(T)/AVQZ, CCSD(T)/AV5Z, and CCSD(T)/AV6Z calculations. (The
SCF energies are of course obtained on the fly.) These
involve basis sets of $[6s5p4d3f2g]$, $[7s6p5d4f3g2h]$, and
$[8s7p6d5f4g3h2i]$ quality, respectively. For the SCF energy, 
we will use the $A+B.C^{-l}$ exponential extrapolation\cite{Fel92}, for the
valence correlation contribution the $A+B/(l+1/2)^C$ formula\cite{l4}.

The extrapolation contributes on average much less than 0.001 eV to the
SCF component, which is basically completely converged with an AV6Z 
basis set. Contributions to the valence correlation energy are somewhat
more significant, reaching -0.016 eV for O and -0.014 eV for F. In all cases,
further basis set expansion is predicted to increase the EA, as expected.

The contribution of inner-shell correlation was obtained by carrying 
out CCSD(T)/ACV5Z calculations both with all electrons correlated,
and with the $(1s)$ like orbitals constrained to be doubly occupied.
While the core correlation energy may converge quite slowly in absolute 
terms, in relative terms (in this case, its contribution to EA)
convergence is usually fairly rapid, with an ACVQZ basis set usually
being large enough even for accurate work, and an ACV5Z basis set
definitely so. As expected, its contribution increases EA in all cases
(except of course for the trivial case of H/H$^{-}$).

The relativistic contribution can be decomposed into two terms: 
the scalar contribution and the effect of  spin-orbit splitting. While
an ab initio purist would obtain the latter from computed spin-orbit coupling 
elements, we have obtained them here from the observed fine structure of
the atomic ground states. Especially for F/F$^{-}$, its contribution to
EA is quite significant; in all cases, a lowering of EA is seen.

The scalar relativistic contribution was obtained by first-order perturbation
theory applied to the Darwin and mass-velocity contributions.\cite{Cow76,Mar83}
Since they can be evaluated as a simple expectation value from the
converged wave function using this method, we have computed them
at the ACPF (augmented coupled pair functional) level.\cite{Gda88}
Relativistic calculations generally require greatly improved flexibility of
the wave function in the high-exponent region, particularly in the $s$
functions: we opted for an uncontracted ACVQZ basis set. In all cases,
the scalar relativistic contribution decreases EA; as expected, the
size of this contribution goes up superlinearly with $Z$. (From a fit
of $a Z^b$ to the computed contributions, we find $b\approx3.4$ in this case.)
While the importance of relativistic contributions to this type of quantity
for heavy elements is well known (e.g., the existence of the alkali
aurides Rb$^+$Au$^-$ and Cs$^+$Au$^-$ is due to relativistic stabilization
of the $6s$ shell\cite{Pyk78}), a contribution of -0.01 eV for an element 
as light a F may seem surprising at first. Since all electron affinities 
discussed here except EA(H) involve addition of an electron to a $2p$
orbital, Kendall et al.\cite{Ken92} conclude that ``relativistic effects
should contribute insignificantly to the calculated electron affinities''.
Of course, whether or not -0.01 eV is insignificant is a matter of the 
accuracy being pursued, as well as the relative magnitude of the other
possible sources of error.

Last but not least, we need to make an allowance for imperfections in
the CCSD(T) treatment. For B/B$^{-}$, we have done so by comparing
an FCI/AVQZ calculation with the corresponding CCSD(T)/AVQZ results.
This calculation took about five hours on an SGI Origin 2000 minisupercomputer,
and could not be carried to completion even for C$^-$. Since however the 
FCI-CCSD(T) correction for EA(B) appears to converge very rapidly
(with the AVDZ, AVTZ, and AVQZ basis sets we obtain values of
0.0186, 0.0197, and 0.0191 eV, respectively), we may fairly safely
use the difference between FCI/AVTZ and CCSD(T)/AVTZ in the case
of EA(C). For O/O$^-$ and F/F$^-$, even FCI/AVTZ calculations are not
feasible. Assuming that the error in the full CCSDT method
with respect to FCI is fairly constant, we have therefore employed a two-stage 
additivity approximation in these cases:
\begin{eqnarray}
E[{\rm FCI/AV\infty Z}]\approx E[{\rm CCSD(T)/AV\infty Z}]&&\nonumber\\
+ \left(E[{\rm CCSDT/AVQZ}] - E[{\rm CCSD(T)/AVQZ}]\right)\nonumber\\
+ \left(E[{\rm FCI/AVDZ}] - E[{\rm CCSDT/AVDZ}]\right)
\end{eqnarray}

As seen in Table \ref{ea}, the final results agree with experiment to
within about 0.001 eV on average. The largest discrepancies, 0.0015 eV,
occur for EA(O) and EA(F), which are also the largest and for which some of the
individual contributions (e.g. the relativistics) are also the largest.

\subsection{Atomization energy of SiH$_4$ and the
heat of formation of Si($g$)}

The heat of formation of Si($g$) is the subject of some controversy.
In the JANAF tables it is given as 106.6$\pm$1.9 kcal/mol. Desai\cite{Desai}
reviewed the available data and recommended the JANAF value, but with a
reduced uncertainty of $\pm$1.0 kcal/mol. Recently, Grev and 
Schaefer\cite{Gre92} found that their ab initio calculation of the TAE of
SiH$_4$, despite basis set incompleteness, was actually {\em larger} than the
value derived from the experimental heats of formation of Si($g$), H($g$),
and SiH$_4$($g$). They suggested that the
heat of vaporization of silicon be revised upwards 
to $\Delta H^\circ_{f,0}$[Si($g$)]=108.07(50) kcal/mol, a suggestion
supported by Ochterski et al.\cite{Och95}. Clearly, some calibration
calculation to resolve this controversy would be desirable: we will
here report some preliminary results obtained as a by-product of
an anharmonic force field study\cite{sih4} on SiH$_4$. While Grev and
Schaefer's work was definitely state of the art in its time, the attainable
accuracy for this type of compound may well have gone up an order of 
magnitude in the six years since it was published.

From a calibration calculation along the lines discussed above, our best
calculation for the nonrelativistic valence CCSD(T) limit is 324.62
kcal/mol. For this molecule, we may assume fairly safely that CCSD(T)
is close to full CI. Deducting the Si spin-orbit splitting correction
(0.43 kcal/mol), adding a core correlation contribution of -0.34
kcal/mol (with the MT core correlation basis set) and deducting a fully
anharmonic zero-point energy of 19.57 kcal/mol (from a CCSD(T)/VQZ+1
quartic force field) we obtain TAE$_0$=304.28 kcal/mol. Using the revised
$\Delta H^\circ_{f,0}$[Si($g$)]=108.07(50) kcal/mol of Grev and Schaefer
\cite{Gre92} we obtain $\Delta H^\circ_{f,0}$[SiH$_4$($g$)]=10.34 kcal/mol,
in excellent agreement with the JANAF value of 10.5(5) kcal/mol. At first
sight this supports the new value.

Upon introducing the scalar relativistic correction of -0.67 kcal/mol, however,
we obtain a value of 11.0 kcal/mol, which is only just compatible with
the experimental measurement. Using the older JANAF/CODATA value 
$\Delta H^\circ_{f,0}$[Si($g$)]=106.6$\pm$1.0 kcal/mol, we would find 9.54 
kcal/mol, seemingly incompatible with the experimental result for SiH$_4$.
However, as pointed out by Grev and Schaefer\cite{Gre92}, the JANAF value
is in fact the Gunn and Green\cite{Gun61} value of 9.5 kcal/mol increased
by a correction\cite{Ros52} of +1 kcal/mol for the phase transition 
Si(amorphous)$\rightarrow$Si(cr). If one were to follow Gunn and Green
in considering this correction to be an artifact of the method of preparation
and in neglecting it, our calculations would in fact support the {\em old}
JANAF/CODATA $\Delta H^\circ_{f,0}$[Si($g$)]. 

Regardless of who is `right' here (CODATA or Grev and Schaefer), 
the above serves as an illustration that, where the accurate determination
of fundamental thermochemical quantities is at stake, the greatest care
is required, both in performing the calculations and in interpreting the
experiments.

\subsection{Heat of formation of B(g) via the total atomization
energy of BF$_3$}

Nonthermochemists are often surprised when they hear that the heats of
formation in the gas phase of three first-and second-row atoms 
(namely, Be, B, and Si) are imprecisely known because of various
experimental complications. The most uncertain
value among them, B, carries an error bar of no less than 3 kcal/mol, 
$\Delta H^0_f$(B($g$))=132.7$\pm$3.0 kcal/mol \cite{Jan85}.
This is obviously a very unsatisfactory state of affairs 
given the fact that just about any ab initio or semiempirical scheme
for calculating molecular heats of formation relies on the heats of formation
of the constituent atoms through the identity
\begin{eqnarray}
\Delta H^\circ_{f,T}(\hbox{A$_k$B$_l$C$_m$\ldots})
   &-& k \Delta H^\circ_{f,T}(\hbox{A})
   - l \Delta H^\circ_{f,T}(\hbox{B})
   - m \Delta H^\circ_{f,T}(\hbox{C}) - \ldots\nonumber\\
=   {\rm TAE}_0 &+& E_T(\hbox{A$_k$B$_l$C$_m$\ldots}) 
   - k E_T(\hbox{A}) - l E_T(\hbox{B})  - \ldots \nonumber\\
   &-& m  E_T(\hbox{C}) + RT (1-k-l-m-\ldots)
\end{eqnarray}
(where T is the temperature).

Storms and Mueller (SM)\cite{Sto77} had previously recommended a much higher
and more precise value of 136.2$\pm$0.2 kcal/mol. Ru\v{s}\v{c}i\'c et al.\cite{Rus88},
reviewing the experimental data, concluded that the JANAF value was in
error and recommended the SM value. Recently, Ochterski et al.\cite{Och95}
combined calculated atomization energies using the CBS-APNO hybrid
ab initio/empirical scheme\cite{Mon94} with an accurate CODATA\cite{Cod89}
heat of formation for BF$_3$, 271.2$\pm$0.2 kcal/mol, and the established
heat of formation for F($g$), 18.47$\pm$0.07 kcal/mol, to obtain 135.7
kcal/mol.
On the basis thereof, they too recommended the SM value.
Note that their calculation does not include a correction for the spin-orbit
splitting in atomic fluorine and therefore is about 1.1 kcal/mol too high (see
below). In another study, Schlegel and Harris\cite{Sch94} found that
computed heats of formation using the Gaussian-2 (G2) method\cite{Cur91}
for a number of boron compounds agreed much better with experiment if
the reference value for gaseous boron was taken as the SM rather than the
JANAF value.

Martin and Taylor\cite{bf3} carried out a calibration calculation aimed 
at resolving this discrepancy for once and for all. All relevant energies
are given in Tables \ref{bf3stuff1} and \ref{bf3stuff2}. 

The largest calculation we could carry out on BF$_3$ was 
CCSD(T)/AV5Z (508 basis functions), which required
60~GB of disk space and 720~MB of memory on the CRAY~T90. Because the
next step up in basis set, CCSD(T)/AV6Z (756 basis functions) was
simply beyond the available computational hardware, we used the
BF diatomic as a model system for the effect of further basis set
extension.

The SCF component of the atomization energy of BF$_3$ differs only
-0.02 kcal/mol between AVQZ and AV5Z basis sets, and is essentially
converged. For BF, increasing the basis set another step to AV6Z only
affects the result by 0.01 kcal/mol; upon exponential extrapolation,
the Feller(Q56) total SCF energy, $-$124.168760 $E_h$, is found to be
only 20 $\mu E_h$ above the numerical Hartree-Fock result\cite{Kob95}.

\begin{table}[h]
\caption{Convergence of individual contributions to the
TAE of BF$_3$ and to $D_e$(BF). All values are in kcal/mol\label{bf3stuff1}}

\begin{tabular}{lccl}
\hline
 & BF$_3$ & BF & ~~~~~~~~~~~~~~~~~~~~~~\\
\hline
\multicolumn{3}{c}{SCF component of TAE$_e$}\\
\hline
SCF/AVTZ & 373.59 & 142.30 \\
SCF/AVQZ & 374.61 & 143.03 \\
SCF/AV5Z & 374.59 & 143.08 \\
SCF/AV6Z & ---    & 143.09 \\
Feller(TQ5) & 374.59 & 143.08$_5$ \\
Feller(Q56) &  ---   & 143.08$_7$ \\
Best SCF$^a$ & 374.59 & 143.09 \\
\hline
\multicolumn{3}{c}{valence correlation component of TAE$_e$}\\
\hline
CCSD(T)/AVTZ & 87.38 & 35.63 \\
CCSD(T)/AVQZ & 91.83 & 37.63 \\
CCSD(T)/AV5Z & 93.19 & 38.19 \\
CCSD(T)/AV6Z & ---   & 38.44 \\
Schwartz$\alpha$(TQ5) & 94.03 & 38.35 \\
Schwartz$\alpha$(Q56) & --- & 38.76 \\
Best valence corr.$^b$ & 95.13 & 38.76 \\
\hline
\multicolumn{3}{c}{Inner shell correlation component of TAE$_e$}\\
\hline
CCSD(T)/CVTZ & 1.366 & 0.482 \\
CCSD(T)/CVQZ & 1.724 & 0.629 \\
CCSD(T)/CV5Z & ---   & 0.670 \\
Schwartz$\alpha$(TQ5) & --- & 0.696 \\
CCSD(T)/ACVTZ & 1.563 & 0.557 \\
CCSD(T)/ACVQZ & 1.772 & 0.648 \\
CCSD(T)/ACV5Z & ---   & 0.676 \\
aug-Schwartz$\alpha$(TQ5) & --- & 0.698 \\
Best core corr.$^c$ & 1.922 & 0.698 \\
\hline
\end{tabular}

(a)
Feller(TQ5)[BF$_3$]$+3\times\left(\hbox{Feller(Q56)[BF]$-$Feller(TQ5)[BF]}\right)$

(b)
Schwartz$\alpha$(TQ5)[BF$_3$]$+
3\times\left(\hbox{Schwartz$\alpha$(Q56)[BF]$-$Schwartz$\alpha$(TQ5)[BF]}\right)$

(c)
CCSD(T)/ACVQZ[BF$_3$]$+3\times\left(\hbox{aug-Schwartz$\alpha$(TQ5)[BF]$-$CCSD(T)/ACVQZ[BF]}\right)$

\end{table}

Improving the basis set from AVQZ to AV5Z increases the valence 
correlation energy
by some 1.39 kcal/mol, compared to 4.46 kcal/mol from AVTZ to AVQZ.
The Schwartz$\alpha$(TQ5) extrapolation adds on another 0.84 kcal/mol; note that
while the value of $\alpha$ for BF$_3$ is about 3.40, 
the $\alpha$ found for the MP2 correlation energy, 2.88, strongly
suggests dominance of the leading $(l+1/2)^{-3}$ term. The
difference between the MP2 and CCSD(T) values of $\alpha$ suggests the
importance of higher-order contributions, which add\cite{Kut92} higher powers
in $(l+1/2)$. For the BF model system, the Schwartz$\alpha$(TQ5) extrapolated 
value is no less than 0.37 kcal/mol below the Schwartz$\alpha$(Q56) value:
this unusually large difference is to some extent due to the very polar 
character of the B--F bond. 
(In fact, Gillespie\cite{Gil98}
argues that BF$_3$ is best regarded as a tricoordinate ionic compound
of B$^{3+}$.)

Since BF$_3$ actually contains three bonds that are quite similar to
the one in BF, it seems reasonable that the difference between
Schwartz$\alpha$(TQ5) and Schwartz$\alpha$(Q56) would be approximately
three times that in BF. Hence we obtain an estimated basis set limit for
the correlation part of TAE of 95.14 kcal/mol. In combination with the
SCF contribution of 374.57 kcal/mol this yields a valence-only TAE,
without spin-orbit correction, of 469.71 kcal/mol.

The contribution of inner-shell correlation to the TAE of BF$_3$ is
found to be 1.37 kcal/mol at the CCSD(T)/CVTZ level and 1.72 kcal/mol
at the CCSD(T)/CVQZ level. Given the polarity of the system, some mild
coupling between the effects of core correlation and inclusion of diffuse
functions cannot be ruled out a priori, and indeed extending the
CVQZ to an ACVQZ basis set adds some 0.05 kcal/mol to the core correlation
energy. Based on experience\cite{cc} we normally expect the
core correlation contribution to be near convergence with such basis sets.

Again using the BF diatomic as a model system permits us to gauge the
effects of further improvement of the core correlation basis set. At the
CCSD(T)/ACVQZ level, the core correlation contribution to $D_e$(BF) is
0.65 kcal/mol, or slightly more than one-third the value in BF$_3$.
Enlarging the basis from CVQZ to CV5Z leads to an increase of
0.04 kcal/mol: the effect from ACVQZ to ACV5Z is somewhat smaller at 0.03
kcal/mol. (The CV5Z and ACV5Z values differ by only 0.01 kcal/mol.)
Carrying out a Schwartz$\alpha$(TQ5) extrapolation on the ACVTZ, ACVQZ, and
ACV5Z numbers leads to an estimated infinite-basis limit core correlation
contribution to the BF $D_e$ of 0.70 kcal/mol, or 0.05 kcal/mol more than
the computed ACVQZ value.

If we again use three times this value as a correction for BF$_3$, we obtain
a best estimate for the inner-shell correlation contribution to TAE(BF$_3$)
of 1.92 kcal/mol. We hence obtain a TAE$_{e,NR}$ (i.e. without spin-orbit
correction) of 471.65 kcal/mol; deducting the atomic spin-orbit corrections
finally yields TAE$_e$=470.46 kcal/mol.

From the computed CCSD(T)/VTZ harmonic frequencies and anharmonicity
constants given in Ref.\cite{Pak97}, we obtain ZPE=7.89 kcal/mol. If we
substitute experimental fundamentals (see Ref\cite{Pak97} for details)
and employ the computed
anharmonicity constants only for the small difference between the
zero-point energy and one-half the sum of the fundamentals, ZPE decreases
to 7.83 kcal/mol. We hence obtain the total atomization energy for BF$_3$
at 0~K, TAE$_0$=462.63 kcal/mol.

\begin{table}
\caption{Computed thermochemical properties for BF$_3$, BF, and B
in the gas phase. All values are in kcal/mol\label{bf3stuff2}}
\begin{tabular}{lccl}
\hline
 & BF$_3$ & BF & ~~~~~\\
 \hline
Best TAE$_{e,NR}$ & 471.65 & 182.54 \\
spin-orbit correction$^a$ & $-$1.184 & $-$0.414 \\
Best TAE$_e$ & 470.46 & 182.13 \\
ZPVE & 7.887$^b$ & 1.996$^c$ \\
Best TAE$_0$ & 462.63 & 180.13 \\
\hline
\multicolumn{3}{c}{Derivation of $\Delta H^0_{f,0}$[B($g$)]}\\
\hline
$\Delta H^0_{f,0}$[BF$_3$($g$)], Ref.\cite{Cod89}$^*$ & $-$270.8$\pm$0.2 \\
$\Delta H^0_{f,0}$[F($g$)], Ref.\cite{Jan85} & +18.47$\pm$0.07 \\
calculated $\Delta H^0_f$[B($g$)] & 136.4$\pm$0.4\\
Expt. JANAF\cite{Jan85} & 133$\pm$3 \\
Expt. SM\cite{Sto77}, 298 K & 137.4$\pm$0.2\\
$\Delta H^0_{f,298}$-$\Delta H^0_{f,0}$, Ref.\cite{Jan85} & 1.219 \\
Expt. SM\cite{Sto77}, 0 K & 136.2$\pm$0.2\\
\hline
\end{tabular}

(*) $\Delta H^0_{f,298}$[BF$_3$($g$)]=271.5$\pm$0.2 kcal/mol; 
$\Delta H^0_{f,0}$[BF$_3$($g$)]-$\Delta H^0_{f,298}$[BF$_3$($g$)]
= -$(H_{298}-H_{0})$[BF$_3$($g$)-B($g$)-3/2 F$_2$($g$)] 
= -(11.65 - 1.222 - (3/2) 8.825)/4.184 = +0.675 kcal/mol. (All data from
Ref.\cite{Cod89}.)

(a) computed from atomic sublevels for electronic ground states given
in Ref.\cite{Jan85}.

(b) from observed $\nu_i$ and computed $X_{ij},G_{ij}$ given in
Ref.\cite{Pak97}

(c) from computed CCSD(T)/VQZ $\omega_e$=1398.0, $\omega_ex_e$=11.55,
and $\omega_ey_e$=0.054 cm$^{-1}$:
experimental values\cite{Hub79} 1402.1$_3$, 11.8$_4$, and 0.05$_6$ cm$^{-1}$,
respectively.

\end{table}

In combination with the JANAF\cite{Jan85} heat of formation for F(g) of
18.47$\pm$0.07 kcal/mol and the CODATA\cite{Cod89} heat of formation
of BF$_3$($g$), $-$270.84$\pm$0.2 kcal/mol, we then obtain
$\Delta H^0_{f,0}$(B($g$))=136.38$\pm$0.3 kcal/mol, in which the uncertainty
only reflects the uncertainties in the experimental quantities. The possible
further error in the calculations is somewhat more difficult to quantify:
past experience suggests a mean absolute error of 0.12 kcal/mol, but in the
light of the fairly substantial correction terms applied, it would probably
be appropriate to increase the error margin to about 0.3 kcal/mol. This would
then bring our best estimate to 136.4$\pm$0.4 kcal/mol, the uncertainty
of which encompasses that of the SM value of 136.2$\pm$0.2
kcal/mol.

In the published study\cite{bf3}, we did not consider two contributions: 
imperfections
in the CCSD(T) method and scalar relativistic contributions. The former
are rather hard to quantify since a full CI calculation for this 
system, even in a fairly small basis set, is not a realistic option at
present. We can determine the latter by an ACPF/CVTZ(uncontracted) 
calculation of 
the Darwin and mass-velocity contributions, which we find to be -0.68 
kcal/mol. Adding another 0.1 kcal/mol to the error bar in order
to accommodate uncertainty in this contribution, we then have a best
estimate for the heat of atomization at 0 K of B of 135.7$\pm$0.5 kcal/mol, 
which is still compatible with the SM value.

\section{Conclusions}

We have shown that
by judicious use of extrapolations to the 1-particle basis set
limit and $n$-particle calibration techniques, total atomization
energies of molecules with up to four heavy atoms can be obtained
with calibration accuracy (1 kJ/mol or better, on average) without
any empirical correction. For the SCF energy a 3-point 
geometric extrapolation is the method of choice. For the MP2 correlation
energy, a 2-point $A+B/(l+1/2)^3$ extrapolation is recommended, while
for CCSD and CCSD(T) correlation energies we prefer the 3-point
$A+B/(l+1/2)^C$ formula. Addition of high-exponent `inner polarization
functions' to second-row atoms is essential for reliable results. 
For the highest accuracy, accounts are required of inner-shell correlation,
atomic spin-orbit splitting, anharmonicity in the zero-point energy, and
scalar relativistic effects.

\section*{Acknowledgments}

The author is the Incumbent of the Helen and
Milton A. Kimmelman Career Development Chair and a Yigal Allon Fellow, 
as well as an Honorary Research Associate
(``Onderzoeksleider in eremandaat'') of the National Science Foundation
of Belgium (NFWO/FNRS). He acknowledges helpful discussions with
many colleagues, especially Drs. Charles W. Bauschlicher Jr. and Timothy J. Lee 
(NASA Ames Research Center), Dr. Thom H. Dunning Jr. (Pacific Northwest
National Laboratories), Prof. Trygve U. Helgaker (Oslo University, Norway),
Prof. Peter R. Taylor (San Diego Supercomputer Center and University of California,
San Diego). Finally, he would like to thank Dr. Kim Baldridge (San Diego
Supercomputer Center) for critical reading of the manuscript prior to 
submission.

Previously unpublished calculations reported in this work were carried
out using ACES II\cite{aces}, Gaussian 94\cite{g94}, 
and a prerelease version of MOLPRO97.3\cite{molpro} 
made available courtesy of Prof. Peter J. Knowles (Birmingham University,
UK). Some unpublished correlation consistent basis sets were 
taken from the EMSL library\cite{EMSL}; others (particularly the AV6Z sets used)
are `unofficial' ones generated by the author.

This research was partially supported by the Minerva Foundation, Munich, Germany,
and by grants of computer time from San Diego Supercomputer Center.

\end{document}